\documentclass[a4paper,11pt]{article}
\usepackage{setspace}
\usepackage[english]{babel}
\usepackage{amsmath}
\usepackage{amssymb}
\usepackage{graphicx}
\usepackage[hidelinks]{hyperref} %[hidelinks] for no ugly boxes
\usepackage{hyperref}
\usepackage{siunitx}
\usepackage{xcolor}
\usepackage{geometry}
\usepackage[affil-it]{authblk}
\usepackage{ulem} % st for strikethrough
\usepackage{placeins} % for FloatBarrier
\usepackage{makecell} % für Zeilenumbruch in Tabellen
\usepackage{geometry}
\usepackage[T1]{fontenc}
\usepackage{braket}
\newcommand{\grayScale}{0.95} % Can change the gray level here
\definecolor{codeBackground}{rgb}{\grayScale ,\grayScale ,\grayScale}
\definecolor{forestGreen}{rgb}{0.13,0.55,0.13}

\graphicspath{{./figures/}}

\newgeometry{left=3cm,right=3cm,top=2cm,bottom=2cm}

\definecolor{Darkgreen}{rgb}{0,0.4,0}
\definecolor{Purple}{rgb}{0.6, 0.0 0.8}

\newcommand{\Be}{$^9$Be$^+$}

\definecolor{listinggray}{gray}{0.9}
\definecolor{lbcolor}{rgb}{0.9,0.9,0.9}

\newcommand{\kkb}{$\text{K} \cdot k_B$}

\newcommand{\subf}[2]{%
	\setlength{\tabcolsep}{0pt} % space between e.g. a) and figure as small as possible
	{\small\begin{tabular}{cc}
			{\fontfamily{phv}\selectfont 
			#2} & \raisebox{-0.95\height}{ #1} \\
		\end{tabular}
	}%
	\setlength{\tabcolsep}{6pt} % setting back to default value
}

		\newcommand{\approptoinn}[2]{\mathrel{\vcenter{
			\offinterlineskip\halign{\hfil$##$\cr
				#1\propto\cr\noalign{\kern2pt}#1\sim\cr\noalign{\kern-2pt}}}}}
\newcommand{\appropto}{\mathpalette\approptoinn\relax}

\begin{document}

	%\preprint{APS/123-QED}
	
	\renewcommand\Authfont{\fontsize{10}{11}\selectfont}
	\renewcommand\Affilfont{\fontsize{9}{11}\itshape}
	
	\newcommand{\mpik}{1}
	\newcommand{\riken}{2}
		\newcommand{\utexas}{3}
	\newcommand{\jgu}{4}
	\newcommand{\uhanov}{5}
	\newcommand{\ptb}{6}
	\newcommand{\cern}{7}
	\newcommand{\utokyo}{8}
	\newcommand{\gsi}{9}
	\newcommand{\helmholtzMz}{10}

	\author[\mpik]{C. Will\footnote{christian.will@mpi-hd.mpg.de}}
	\author[\mpik, \riken]{M. Bohman\footnote{Present address: High Energy Physics Division, Argonne National Laboratory, 9700 S Cass Ave, Lemont, IL 60439 }}
	\author[\utexas]{T. Driscoll}
	\author[\mpik,\riken]{M. Wiesinger}
	\author[\jgu]{F. Abbass}
	\author[\riken,\uhanov,\ptb]{M. J. Borchert}
	\author[\riken,\cern]{J. A. Devlin}
	\author[\riken,\cern]{S. Erlewein}
	\author[\riken,\utokyo]{M. Fleck}
%	\author[3]{S. Gavranovic}
%	\author[1,2]{J. Harrington}
%	\author[1,2,6]{J. Jaeger}
	\author[\riken]{B. Latacz}
%	\author[2,6]{P. Micke}
	\author[\jgu]{R. Moller}
	\author[\mpik]{A. Mooser}
	\author[\jgu]{D. Popper}
	\author[\mpik,\riken,\cern]{E. Wursten}
	\author[\mpik]{K. Blaum}
	\author[\utokyo]{Y. Matsuda}
	\author[\uhanov,\ptb]{C. Ospelkaus}
	\author[\gsi]{W. Quint}
	\author[\jgu,\helmholtzMz]{J. Walz}
	\author[\riken,\jgu]{C. Smorra}
	\author[\riken]{S. Ulmer}

	\affil[\mpik]{Max-Planck-Institut für Kernphysik, Saupfercheckweg 1, D-69117 Heidelberg, Germany}
	\affil[\riken]{RIKEN, Fundamental Symmetries Laboratory, 2-1 Hirosawa, Wako, Saitama 351-0198, Japan}
		\affil[\utexas]{Department of Physics, The University of Texas at Austin, Austin, Texas 78712, USA}
	\affil[\jgu]{Institut für Physik, Johannes Gutenberg-Universität, Staudingerweg 7, D-55128 Mainz, Germany}
	\affil[\uhanov]{Institut f{\"u}r Quantenoptik, Leibniz Universität Hannover, D-30167 Hannover, Germany}
	\affil[\ptb]{Physikalisch-Technische Bundesanstalt, D-38116 Braunschweig, Germany}
	\affil[\cern]{CERN, 1211 Geneva, Switzerland}
	\affil[\utokyo]{Graduate School of Arts and Sciences, University of Tokyo, Tokyo 153-8902, Japan}
	\affil[\gsi]{GSI Helmholtzzentrum für Schwerionenforschung GmbH, D-64291 Darmstadt, Germany}
	\affil[\helmholtzMz]{Helmholtz-Institut Mainz, D-55099 Mainz, Germany}

	%\renewcommand*{\Affilfont}{\small\normalfont}
	
	%	\collaboration{BASE Collaboration}%\noaffiliation
	
	\date{\today}% It is always \today, today,
	
	\title{Sympathetic cooling schemes for separately trapped ions coupled via image currents}
	\maketitle

	\begin{abstract} 
		
	Cooling of particles to mK-temperatures is essential for a variety of experiments with trapped charged particles. However, many species of interest lack suitable electronic transitions for direct laser cooling. 
	We study theoretically the remote sympathetic cooling of a single proton with laser-cooled \Be{} in a double-Penning-trap system.
	We investigate three different cooling schemes and find, based on analytical calculations and numerical simulations, that two of them are capable of achieving proton temperatures of about 10\,mK with cooling times on the order of 10\,s. In contrast, established methods such as feedback-enhanced resistive cooling with image-current detectors are limited to about 1\,K in $ 100$\,s. 
	Since the studied techniques are applicable to any trapped charged particle and allow spatial separation between the target ion and the cooling species, they enable a variety of precision measurements based on trapped charged particles to be performed at improved sampling rates and with reduced systematic uncertainties.

	\end{abstract}

	\section{Introduction}
	The ability to prepare (ultra-)cold particles in a controlled environment through laser cooling has been a major driving force in the enhanced precision of recent experiments in the field of atomic, molecular and optical physics  \cite{Saf18}.
	 However, many particle species of interest offer no suitable transitions for direct laser cooling, so that for these particles sympathetic cooling techniques have to be employed \cite{Boh17, Tu21, Bak21, Mad14, Sch20, Bre19}.
	Established sympathetic cooling schemes rely on Coulomb interactions between the target particle and the laser-cooled ions.  
	In order to obtain a sufficiently high coupling rate, the two species must either be co-trapped in the same potential well \cite{Bre19, Mic20} or trapped in separate potential wells \cite{Har11, Bro11} that enable particles distances of $10-\SI{100}{\micro m}$. However, trap designs that enable these length scales and the Coulomb interaction between the particle species impose large systematic shifts on potential precision measurements in Penning traps \cite{Sch19}.

	One coupling technique which is not accompanied by these drawbacks is image-current coupling. Here, the charged particle species are coupled through image currents induced in adjacent trap electrodes, allowing a macroscopic separation of the target species and cooling species.
	Recently, our group has published the first experimental results on sympathetically cooling a single proton in a Penning trap by coupling it to a cloud of laser-cooled \Be{} ions located in a separate trap \cite{Boh21}. 
	In this experiment, the coupling between the ions was mediated by a common superconducting RLC resonator, which is usually used for particle detection \cite{Nag16}. A temperature reduction of the axial mode from $17.0\,(2.4)$\,K to $2.6\,(2.5)$\,K was achieved. Other recent experimental efforts have demonstrated the coupling between two highly charged ions with a comparable experimental setup \cite{Tu21} or the coupling of two laser-cooled ions stored in a Paul trap by a floating room-temperature wire \cite{An21}.
	
	The recent developments \cite{Boh21} will be one of the cornerstones for the next generation of precision measurements at low energies. For example, our collaboration (BASE) probes the fundamental charge, parity and time (CPT) reversal symmetry by performing ultra-high-precision Penning-trap measurements on protons and antiprotons \cite{Smo15}. In particular, their charge-to-mass ratio is compared at a relative precision of \SI{69e-12}{} \cite{Ulm15} and their magnetic moments are measured with fractional resolutions of \SI{3e-10}{} \cite{Sch17} and \SI{1.5e-9}{} \cite{Smo17}, respectively; all being the most precise measurements of these quantities to date.
	However, especially the magnetic moment measurements are limited by the nonzero particle temperature of $\gtrsim 1$\,K.
	Other experimental programs that are subject to comparable limitations include the spectroscopy of highly-charged ions \cite{Mic20, Stu19}, the cooling of antiprotons and positrons for the synthesis of cold antihydrogen \cite{Mad14, Per15, Ams21}, mass measurements of heavy elements \cite{Sch20, Rod12}, or high-precision metrology with atomic clocks \cite{Bre19, Hun16}.

	In this work, we study image-current coupling mechanisms by means of combined analytical calculations and numerical simulations. Since the particles and the RLC resonator constitute a complex stochastic system of three coupled oscillators, we use the numerical simulations to study the dynamical behavior and optimize cooling times and temperature limits. The relevant physics is implemented into the simulation code from first principles and experimentally established models only.
	After demonstrating agreement between our simulation results and existing experimental data, we assess the feasibility and performance of three different cooling schemes. We find that with  two of the techniques, coupling via a common electrode and off-resonant coupling via a common RLC circuit, temperatures of $\approx$ 10\,mK can be reached with cooling time constants on the order of $5-20$\,s. Compared to the established method of resistive cooling with an RLC resonator at $4-20$\,K \cite{SchTh}, this constitutes a temperature reduction by a factor of $\approx 100$ and a cooling time reduction by a factor of $\approx 10$. The third technique, on-resonance coupling with a common RLC  resonator, can provide cooling time constants below 1\,s but is limited to particle temperatures in the hundred mK-regime with the current experimental setup.
	
	Although applicable to many systems, we find the image-current cooling technique to be of immediate relevance to ultra-high precision measurements in Penning traps and therefore motivate the requirements on the cooling performance with a future measurement of the proton magnetic moment.

%%%%%%%%%%%%%%%%%%%%%%%%%%%%%%%%%%%%%%%%%%%%%%%%%%%%%%%%%%
	
	\label{sec:introduction}

	\section{Magnetic moment measurements in a Penning trap}
	\subsection{Penning-trap basics}
	\label{sec:penning_trap_basics}
	
\begin{figure}
		\centering
		\centerline{
			\begin{tabular}[t]{ll}
				\rule{0pt}{-1.0ex}%
%				\imagetop{\subf{\includegraphics[width=0.46\linewidth]{basic_trap_schematic3.png}}{(a) } }  &
				\subf{\includegraphics[width=0.45\linewidth]{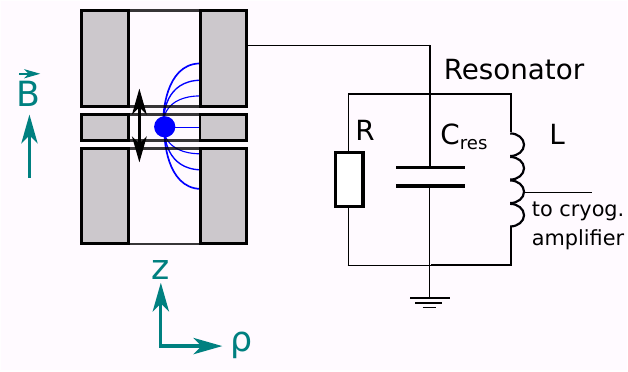}}{(a) }   &
				\subf{\includegraphics[width=0.45\linewidth]{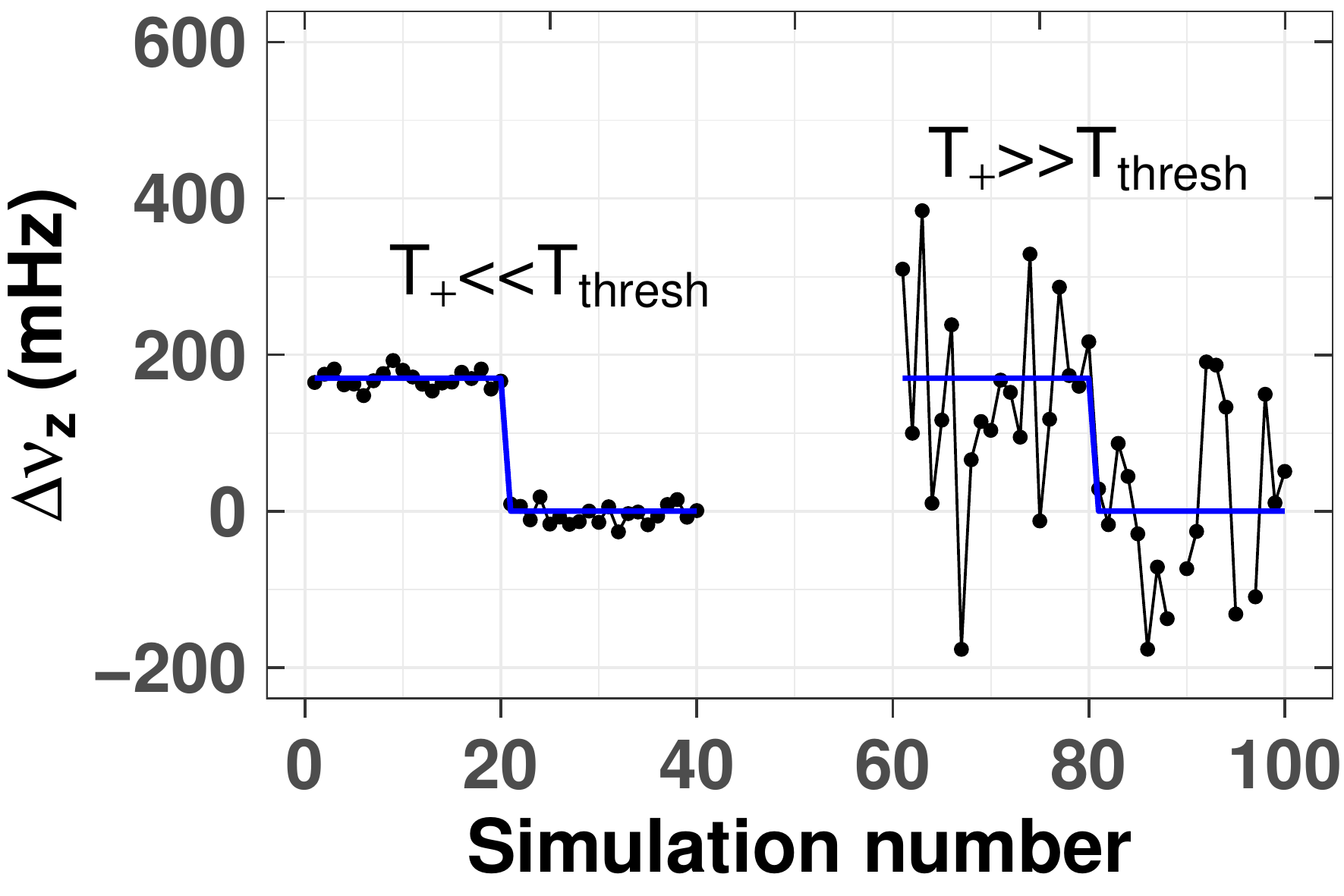}}{ (b) }\\
			\end{tabular}
		}
		\caption{(a) Simplified schematic of a cylindrical Penning trap. The axial motion of the ion induces image currents in the trap electrodes, which are detected by means of a  resonant circuit with high quality factor. 
		(b) Simulated  axial frequency data for low and high modified cyclotron temperatures using the model described in the text. The blue lines indicate the axial frequency without fluctuations from modified cyclotron transitions. A spin flip corresponding to a $\approx 170$\,mHz frequency shift occurs at  simulation number 20 and 80.  The spin flip is easily resolved at a low modified cyclotron temperature. In contrast, at high $T_+$ the superimposed fluctuations prevent an unambiguous identification of a spin flip.  The last magnetic moment measurement of the proton \cite{Sch17} used $T_\text{thresh} = 600$\,mK. }
		\label{fig:trap_schematic_and_cyclotron_quantum_jumps}
	\end{figure}

	In a Penning trap, charged particles are trapped radially by a constant homogeneous magnetic field, in cylindrical coordinates $(\rho, \varphi, z)$ oriented in the axial direction,
	\begin{align}
		\vec{B} = B_0 \, \vec e_z \text{,}
	\end{align}
	 and axially by a harmonic electrostatic potential,
	\begin{align}
		\phi(\rho, z) = C_2 U_0 (z^2 - \rho^2/2) \text{,}
	\end{align}     
	usually created by a stack of five cylindrical electrodes in a compensated, orthogonal design \cite{Gab89}.
	$C_2$ is a geometrical parameter defined by the trap size and $U_0$ the electrical potential difference between the endcap and the central ring elecrode \cite{Bro86}. A simplified schematic is shown in Fig.\,\ref{fig:trap_schematic_and_cyclotron_quantum_jumps}(a). If higher-order contributions to the magnetic field and axial potential are neglected, the axial mode in $z$-direction is independent of the other coordinates and has frequency
	\begin{align}
		\nu_\text{z} = \frac{1}{2\pi} \sqrt{\frac{2 q C_2 U_0}{m}},
	\end{align}
	where  $q$ and $m$ are the charge and the mass of the particle, respectively.
	The radial motional modes split up into a slow and fast mode, referred to as magnetron ($\nu_-$) and modified cyclotron ($\nu_+)$ modes. At $B_0 \approx 2\,$T and $U_0 \approx 1-10$\,V, typical values for the frequencies of the proton in the trap system used in our experiments are $\nu_z \approx 600$\,kHz, $\nu_- \approx 7$\,kHz and $\nu_+ \approx 29$\,MHz.
	
	 The motion of the particle induces image  currents in the trap electrodes that permit a non-destructive measurement of the particle's motional frequencies. 
	Fundamental properties of the trapped particle such as its charge-to-mass ratio and magnetic moment can be derived from these frequency measurements \cite{Bla10}. The axial mode is most commonly detected \cite{Win75}, for which the magnitude of the induced image current is 
\begin{align}
	I_{\text{ind}} = \frac{q}{D} \dot z.
\end{align}
	Here $D$ is a specific length related to the  trap size and the chosen pickup geometry \cite{Win75, Sho38}	and can be considered an ``effective electrode distance''.  $\dot{z}$ is the particle's axial velocity.
	Detectors for these currents consist of ultra low-noise amplifiers connected to superconducting NbTi coils, in our experiment with inductances $L$ of several mH, which are cooled down to \SI{4.2}{K} by a liquid helium bath \cite{Nag16, Ulm09}. Due to parasitic effects, the coil needs to be modelled by an additional  capacitance $C_\text{res}$ and a resistance $R$, forming a parallel RLC circuit with eigenfrequency $\nu_0$ as depicted in Fig.\,\ref{fig:trap_schematic_and_cyclotron_quantum_jumps}(a). The quality factor $Q = R/(2\pi\nu_0 L)$ of such an axial resonator is typically on the order of \SI{20000}{}, including limitations  caused by the input resistance of the cryogenic amplifiers and parasitic resistive losses  \cite{Nag16, Ulm09}. 
	The thermal Johnson-Nyquist noise of the parallel resistance,  $\braket{U_{\text{noise}}^2 } = 4 k_B T_\text{res} R \Delta f$ for a frequency interval $\Delta f$,  charges up the whole RLC circuit and a resonance signal at the circuit's eigenfrequency is created. 
	
	By connecting this resonator to a trap electrode, a trapped  charged particle is exposed to the noisy oscillations of the RLC circuit and  the axial energy dynamically samples a Boltzmann distribution with temperature $T_\text{res}$.  The coupling strength between ion and resonator depends among other parameters on the ion's frequency detuning relative to the resonator. On resonance, where the coupling is strongest, the particle effectively shorts the resonator noise, creating a characteristic dip in the fast Fourier transform (FFT)  spectrum \cite{Win75}. An  example of such a  spectrum is shown in Fig.\,\ref{fig:example_spectrum} of Sec.\,\ref{sec:comparison}. 
	
	In the context of our studies, we will refer to energy as the total oscillator energy, which is, although given in units of \kkb, defined deterministically for every point in time. For example, for the axial mode of a particle, $E_\text{z} = \left( q C_2 U_0 z^2 + \frac{1}{2} m \dot{z}^2 \right)$. In contrast, we will refer to temperature as the average energy of the oscillator divided by the Boltzmann constant, i.e. $T_\text{z} = \braket{E_z}/k_B$.  
	
	\setlength{\tabcolsep}{0pt}
	
		\def\imagetop#1{\vtop{\null\hbox{#1}}}

	\subsection{Magnetic moment measurements and the need for advanced cooling techniques}
	\label{sec:magnetic_moment_measurements}
	Penning-trap based magnetic moment measurements of the proton and antiproton determine the free cyclotron frequency $\nu_\text{c} = \frac{1}{2\pi} \frac{q}{m}B$ and the spin precession frequency (Larmor frequency) $\nu_\text{L} = \frac{1}{2\pi} \frac{g}{2} \frac{q}{m}B$ simultaneously.
	 The ratio of the two quantities is independent of the magnetic field and charge-to-mass ratio of the particle and yields the dimensionless $g$-factor:
	\begin{align}
	g = 2 \frac{\nu_\text{L} }{\nu_\text{c}}.
	\end{align}
	$\nu_\text{c}$ is determined  using the Brown-Gabrielse invariance theorem  $\nu_\text{c} = \sqrt{\nu_+^2 + \nu_\text{z}^2 + \nu_-^2}$ \cite{Bro86}   and measuring $\nu_+$, $\nu_\text{z}$, $\nu_-$  via image current techniques and sideband coupling \cite{Cor90} or phase-sensitive detection \cite{Stu11}. However,  
	 a direct measurement of the Larmor frequency is not possible as the spin orientation cannot be detected by image currents in the trap electrodes. Instead,
	spin flips are detected by utilizing the continuous Stern-Gerlach effect \cite{Smo15}.  Here, a strong magnetic inhomogenity is superimposed to the homogeneous field, $B_\text{z}(\rho, z) = B_0 + B_2 (z^2 - \rho^2/2)$, where $B_2$ characterizes the strength of the magnetic inhomogeneity. This results in the axial frequency becoming a function of both the energy of the radial modes as well as  the spin state of the trapped particle. 
	Spin flips can be induced by applying an external rf drive with frequency close to $\nu_\text{L}$. Measuring the spin-flip probability as a function of the drive frequency yields a resonance line from which $\nu_\text{L}$ can be extracted \cite{Bro84}. However, due to frequency shifts caused by the  nonzero mode temperatures, the strong magnetic bottle  poses limitations to the measurement precision of $\nu_\text{c}$ and $\nu_\text{L}$.  This limitation can be overcome by employing multi-trap techniques \cite{Haef03, Moo13_PhysLettB, Moo14}: The  frequency measurements for the determination of $\nu_\text{c}$ and $\nu_\text{L}$ are conducted in the precision trap (PT) with a homogeneous magnetic field, while the spin state  is determined in the analysis trap (AT)  featuring the magnetic bottle. In the AT, despite using a strong $B_2 = \SI{300 000}{T/m^2}$, a spin flip of a single proton corresponds to an axial frequency shift of only $\approx 170$\,mHz compared to $\nu_\text{z} \approx 600$\,kHz for our experimental parameters. As a result, the axial frequency stability is a key parameter for high-fidelity identification of the spin-state \cite{Moo13_PhysLettB, Smo17_PhysLettB}. 
	We observe that a major contribution of axial frequency fluctuations can be attributed to modified cyclotron quantum transitions \cite{Moo13_PhysLettB}. The  transition rate $\zeta_+$ is proportional to the mode energy:
	\begin{align}
		\zeta_+ \propto n_+ \propto E_+.
	\end{align}
	Here $n_+$ denotes the quantum number of the cyclotron  mode with energy  $E_+ = h \, \nu_+ (n_+ + 1/2) = k_B T_+ $.  The effect of noise-driven cyclotron transitions on a sequence of axial frequency measurements or simulations is illustrated in Fig.\,\ref{fig:trap_schematic_and_cyclotron_quantum_jumps}(b) for the two cases of low and high cyclotron temperatures.  At low $T_+$, the spin flip is easily resolved from residual fluctuations. In contrast, at high $T_+$ the superimposed fluctuations prevent  a high-fidelity identification of a spin flip. Since the measurement of $\nu_+$ increases the modified cyclotron energy,  the preparation of the cyclotron mode with a subthermal energy must be conducted in each double-trap measurement cycle. 
	
	Our latest magnetic moment measurements employed the  technique of resistive feedback cooling with a cyclotron resonator \cite{Dur05, Ulm13}. 
	 When coupled to the resonator, the cyclotron mode thermalizes with the resonator and samples a Boltzmann distribution, in the case of the proton measurement with an average temperature of 3\,K \cite{SchTh}. However, at the same time the acceptance threshold $T_\text{thresh}$ for unambiguous spin-state determination was set to $T_+ \leq T_\text{thresh} = 600$\,mK, so that a time-consuming stochastic selection of the cyclotron energy was necessary. In fact, $2/3$ of the  duration of  one full experimental cycle of about 90\,min were needed for the preparation of  the particle in a state of low cyclotron energy and the subsequent spin-state detection \cite{Sch17}.
	As a consequence, the measurement of the proton $g$-factor is limited by the statistical uncertainty, and the  development of cooling schemes that achieve colder particle temperatures in a shorter time is crucial for overcoming the current limit in precision.
	Ultimately, more powerful cooling methods would not only enable an enhanced spin state detection fidelity, but also allow to deterministically prepare the proton with a cold cyclotron mode in much shorter times, which is one of the cornerstones for achieving $< 10^{-10}$ fractional precision in the next generation of (anti-)proton magnetic moment measurements.

	In order to realize this boost in precision experimentally, we develop technologies and methods to sympathetically cool the proton with a cloud of   laser-cooled  \Be stored in a separate trap a few cm  away, for which we have recently performed the proof-of-principle measurement \cite{Boh21}. The coupling is achieved through image currents induced in a common electrode or a common resonator.
	Since image-current coupling requires matching  an oscillation frequency of the two species involved and the  cyclotron and magnetron frequencies are, in contrast to the axial frequency, not easily adjustable in our experiment, we currently apply the sympathetic cooling technique to the axial mode. 
	The subsequent coupling between the cyclotron mode and axial mode is usually performed via a quadrupolar radiofrequency drive (sideband coupling)  \cite{Cor90}. However, with this technique the mode temperatures are related by $T_+ = \frac{\nu_+}{\nu_\text{z}} T_\text{z} $, in our case $T_+ \approx 60\,T_\text{z}$ for the proton. 
	In order to achieve a significantly improved cyclotron mode preparation at the previous threshold of $T_+ \leq 600$\,mK, cooling techniques that reach axial proton temperatures of $T_\text{z} \lesssim 10$\,mK are required. 
	For example, $T_\text{z} =  10$\,mK axial temperature would  enhance the success rate of the cyclotron energy selection from $\approx 18$\% to $\approx 63$\%. Furthermore, the temperature distribution after the $\leq 600$\,mK-selection is shifted towards lower temperatures, leading to an improved spin-state detection fidelity and a reduction in time for determining the spin state.
%	integral exp(-T/(kb*Tdet))/(kb*Tdet) dT from 0 to Tthresh = 1 - exp(-Tthresh/(kb*Tdet)) -> Tdet=3K vs. Tdet=600mK.
	Since the main objective is to reduce the experimental cycle time and the cyclotron mode needs to be prepared for each measurement cycle, the cooling scheme itself should require less than one minute. 
	Apart from the cooling performance, the technique must  be applicable to both protons and antiprotons and should not induce relative systematic frequency shifts larger than $1\times10^{-11}$.

	\section{Sympathetic cooling techniques for a single proton}
	\label{sec:coupling_mechanisms}

\begin{figure}
	\centering
	\centerline{
		\begin{tabular}[tb]{ll}
			\rule{0pt}{-1.0ex}%
			\imagetop{\subf{\includegraphics[scale=0.98]{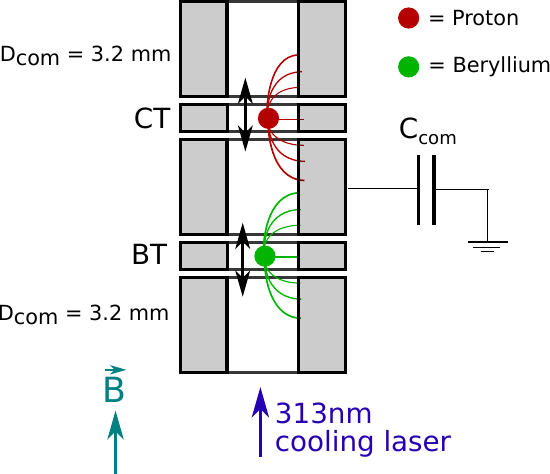}}{(a) } }  &
			\imagetop{\subf{\includegraphics[scale=0.98]{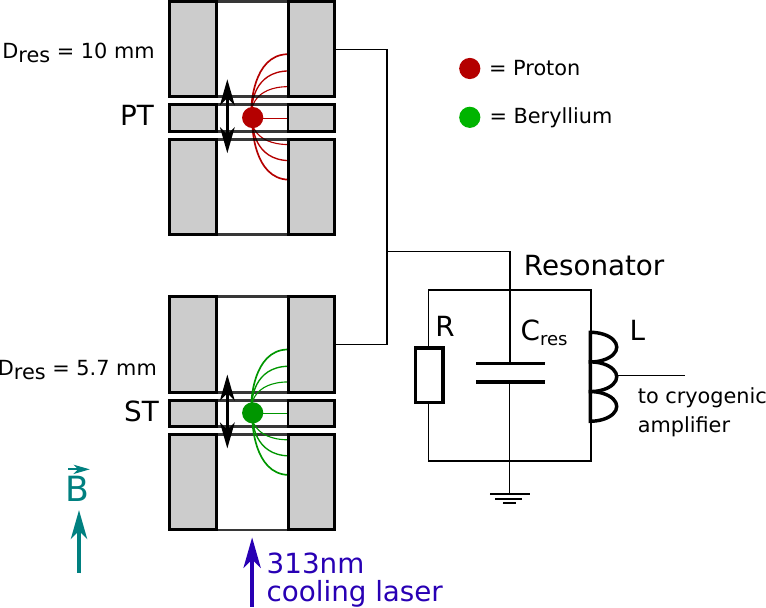}}{ (b) } }\\
		\end{tabular}
	}
	\caption{(a) Sketch of the experimental setup for the common-endcap coupling. The common endcap electrode can be modelled as a  ideal capacitor.  (b) Sketch of the experimental setup for the common-resonator coupling. The proton and  \Be{} ions are stored in two independent traps which share a common resonator. }
	\label{fig:experimental_setup_together}
\end{figure}

	One technique which satisfies the challenging requirements described in Sec.~\ref{sec:introduction} and Sec.~\ref{sec:magnetic_moment_measurements}	is image-current coupling, where the proton is sympathetically cooled by a cloud of  laser-cooled  \Be stored in a separate trap and the particles are coupled through image currents.
	A trap stack consisting of five traps was developed for our experiment, of which four are used to implement the coupling schemes and one is used for spin state analysis \cite{Boh21, BohTh,WieTh}. From an instrumentation point of view, our apparatus allows two different types of image-current couplings, namely coupling via a common endcap electrode or coupling via a common RLC resonator. The experimental configurations are depicted schematically in Fig.\,\ref{fig:experimental_setup_together} and described in more detail in Refs.\,\cite{Boh17, Boh21, Hei90}. 

	When performing the common-endcap coupling (Fig.~\ref{fig:experimental_setup_together}(a)), the proton and beryllium ions are stored in the coupling trap (CT) and beryllium trap (BT), respectively. These traps are designed such that they share a common endcap with which the coupling scheme in the proposals of Refs.~\cite{Boh17, Hei90} can be realized. The image current induced by the particles charges up the capacitance of the common-endcap electrode. The resulting voltage generates a force on the particles, so that a system of two coupled oscillators is created. 
	By laser cooling the beryllium ions with the  313\,nm $^2\text{S}_{1/2} \rightarrow {^2\text{P}_{3/2}}$ transition (natural linewidth: $ \approx 2\pi \times \SI{20}{MHz}$), they act as a heat sink and the proton will be sympathetically cooled as well. This specific approach has the advantage that no additional source of heating is present. Alternatively, we can trap the particles in the precision trap (PT) and the source trap (ST), which are connected by a common RLC resonator, as depicted in Fig.\,\ref{fig:experimental_setup_together}(b). Since the resonator is an oscillator itself, this  configuration represents a system of three coupled oscillators. The advantage of the RLC resonator is that it amplifies the image current signal by the $Q$-factor of the resonator, resulting in a fast energy exchange between the particles. In turn, however, the cooling effect competes with the stochastic heating  by the noise, ultimately limiting  the minimal proton temperature. This resonator-coupling scheme was employed in the  first demonstration of sympathetic cooling through image currents recently conducted by our group \cite{Boh21}.
	The particle-to-detector coupling rate, which governs both the heating and cooling rate, can be reduced by detuning the particles from the resonator eigenfrequency. The reduction in heating rate allows for colder particle temperatures and this specific scheme will be numerically investigated in Sec.\,\ref{sec:off_resonant_coupling}.

	\section{Numerical implementation}
	\label{sec:numerical_implementation}
	
	Next, we describe the numerical implementation of the equations governing our simulations.
	 The proton and  \Be{} ions induce image currents \cite{Win75}
	\begin{equation}
		I_{\text{res/com,p}} = \frac{q}{D_\text{{res/com,p}}} \dot z_\text{p} \qquad \text{and} \qquad I_\text{res/com,Be} = \frac{q}{D_\text{res/com,Be}} \dot z_\text{Be}
	\end{equation}
	in the resonator or the common endcap electrode, denoted with subscript ``res'' and ``com'', respectively.
	In all cases $D_\text{res/com,p/Be}$ is a geometrical parameter defined by the trap geometry and can be considered an ``effective electrode distance''. The common endcap is modelled as an ideal capacitor $C_\text{com}$ charged by the image currents. The potential $U_\text{com}$ due to the accumulated charge leads to a force on the particles, $F_{\text{com,p/Be}} = \frac{q}{D_{\text{com, p/Be}}} U_\text{com}$.
	Similarly, the particles are driven and damped by the oscillating voltage of the resonator $U_\text{RLC}$ at a corresponding force $F_\text{RLC} = \frac{q}{D_\text{res,p/Be}} U_\text{RLC} =\frac{q}{D_\text{res,p/Be}} L \, \dot I_\text{L}$. The resonator itself  is described by a second-order differential equation. In addition to the ion-induced currents, the resonator is charged by the effective thermal Johnson-Nyquist noise current $I_\text{noise}$. 
	We  describe the resonator in terms of the current through the coil $I_\text{L}$. The full set of coupled differential equations, expressed in terms of input parameters of the simulation, is then: 
	%\begin{equation}
	\begin{align}
		\label{eq:finalset1a}
		m_\text{p} \, \ddot z_\text{p} + 2 \, q_\text{p} \, C_\text{2,p} \, U_{0,p} \, z_\text{p}  - \frac{q_\text{p}}{D_\text{res,p}} L \, \dot I_L - \frac{q_\text{p}}{D_\text{com,p}} U_\text{com} &= 0\\
		\label{eq:finalset1b}
		m_{\text{Be}} \,  \ddot z_{\text{Be}} + 2 \, q_\text{Be} \, C_\text{2,Be} \, U_\text{0,Be}\, z_{\text{Be}}  - \frac{q_\text{Be}}{D_\text{res, Be} } L \, \dot I_L + \frac{q_\text{Be}}{D_\text{com,Be}} U_\text{com} &= 0\\
		\label{eq:finalset1c}
		L\, C_\text{res} \, \ddot I_L + \frac{L}{R} \dot I_L + I_L + I_\text{noise} + \frac{q_\text{p}}{D_\text{res,p}} \dot z_\text{p} + \frac{q_\text{Be}}{D_\text{res,Be}} \dot z_{\text{Be}} &= 0 \\
		\label{eq:finalset1d}
		U_\text{com} = \frac{1}{C_\text{com}} \int \left( \frac{q_\text{p}}{D_\text{com,p}} \dot z_\text{p} - \frac{q_\text{Be}}{D_\text{com, Be}} \dot z_{\text{Be}} \right) \text{dt}.
	\end{align}	
	Here, the first two equations describe the axial motion of the particles with external forces $F_\text{RLC}$ and $F_\text{com}$ and the latter two the resonator and common endcap capacitance.
	The subscripts p and Be  denote the quantities related to the proton and  the \Be{} ion(s), respectively.
	
	We neglect the simulation of the radial modes since residual magnetic and electric field inhomogeneities which would couple the axial and radial modes are sufficiently small in our experiment. Furthermore, we neglect Coulomb-interaction between the beryllium ions since only the common mode of the beryllium cloud interacts with the resonator, as we will outline in Sec.\,\ref{sec:resonant_coupling}. 
	
	We solve  Eq.~(\ref{eq:finalset1a})$-$(\ref{eq:finalset1d}) numerically with typical time steps of $\Delta t = 1-5$\,ns compared to $\approx \SI{600}{kHz}$ oscillator frequency.  The time steps are chosen such that  effects which can be attributed to the finite time step are negligibly small and do not affect the coupling studies while still ensuring a reasonable computation time. 
	If the motion of more than one beryllium ion is simulated, either an effective cloud or an array of individual beryllium ions is calculated, depending on whether the single particle behaviour is of interest. 
	A key ingredient of our simulation code is the proper  implementation of $I_\text{noise}$. Eq.\,(\ref{eq:finalset1c})  is a type of Langevin equation where the resistance $R$ determines, according to the fluctuation-dissipation-theorem, both the dissipation and noise terms. In  this case $I_\text{noise}$ obeys a delta-correlated Gaussian probability distribution with zero mean.  The mean squared value of the single-sided power spectral density is given by $\left<I_\text{noise}^2\right> = 4 k_B T_\text{res} \Delta f/R$.  In the time domain representation, the bandwidth $\Delta f$ must be substituted by the Nyquist frequency, which is equal to half of the sampling frequency $f_s = 1/\Delta t$ according to the Nyquist–Shannon sampling theorem, yielding $\Delta f = 1/(2\Delta t)$. Furthermore, in order to achieve a delta-correlation of two subsequent noise samples, each noise value is multiplied with a random value drawn from a Gaussian distribution with a mean of 0 and a standard deviation of 1, abbreviated as $\mathcal{N}_\text{n}(0,1)$.
	In total, in each step $n$ the equation for the noise sample then reads
	\begin{equation}
		I_\text{noise,n}  = \sqrt{2 k_B T_\text{res}/(R \Delta t)} \cdot \mathcal{N}_\text{n}( 0,1).
	\end{equation}
	Each equation in Eq.\,(\ref{eq:finalset1a})-(\ref{eq:finalset1d}) is solved  individually by employing a fourth-order symplectic integrator \cite{Yoshidacite}. This method conserves a Hamiltonian for conservative forces and, although the resonator noise damps and excites the particles, avoids long-term drifts and reproduces energy conservation in the limit $R \rightarrow \infty$. 
	While the integrator can be  applied directly to the  equations of motion of the particles, the integrator for the resonator equation requires a modification due to the dissipative term. 
	Following the notation in Ref.\,\cite{Yoshidacite}, one step for the discretized RLC circuit equation then reads:
	\begin{equation}
		\begin{aligned}
			I_\text{n}^{(i)} &= I_\text{n}^{(i-1)} + c_i \dot I_\text{n}^{(i-1)} \Delta t  \\
			\dot I_\text{n}^{(i)} &= \frac{1}{1+\frac{d_i \Delta t}{RC} } \dot I_\text{n}^{(i-1)} - \frac{1}{1+\frac{d_i \Delta t}{RC} } \frac{d_i \Delta t}{L C} \left( I_\text{n}^{(i-1)} + I_\text{ext} \right).
		\end{aligned}
	\end{equation}

	Here $i \in \{1,2,3,4\}$ denotes the sub-steps where $I_\text{n}^{(0)} \equiv I_\text{n}$, $ \dot I_\text{n}^{(0)} \equiv \dot I_\text{n}$,  $I_\text{n}^{(4)} \equiv I_{n+1} $ and $ \dot I_\text{n}^{(4)} = \dot I_\text{n}^{(3)} \equiv \dot I_{n+1} $. Besides, $I_\text{ext} = I_\text{noise} + I_\text{res,p} + I_\text{res,Be}$ is the external current. 
	In order for the algorithm to be symplectic, the dimensionless coefficients $c_i$ and $d_i$, which define the length of the interleaved time steps, must be chosen appropriately. While two out of four coefficients are negative, i.e. correspond to a step backwards in time, their sum obeys $\sum c_i = \sum d_i = 1$.

	Laser cooling of the \Be{} ions is  modelled by assuming that a narrow linewidth laser with frequency $f_\text{L}$, detuned by $\delta_\text{L}$ relative to the transition frequency $f_0 = f_\text{L} - \delta_\text{L}$, interacts with the  \Be{} ions. 
	In each simulation step in which a  \Be{} ion is in the ground state it can absorb a photon with probability 
	\begin{align}
		P_\text{abs} = B_{12} (\dot{z}_\text{Be}) \Delta t = \frac{I_\text{laser}}{I_\text{sat}} \frac{\Gamma \Delta t / 2}{1 +4 \left(  \frac{2\pi \delta_\text{L} + k_\text{L} \dot{z}_\text{Be} }{\Gamma}\right)^2}.
	\end{align} 
	 $B_{12}$ and $B_{21}$  are the Einstein rate coefficients for absorption and emission, respectively. Further, $k_\text{L}$ is the laser wave vector in axial direction, $\Gamma$ the  natural linewidth of the transition ($\approx 2\pi \times 20$\,MHz in our case), $I_\text{laser}$ the laser intensity and $I_\text{sat} = \frac{\pi  h \Gamma f_\text{0}^3}{3 c^2}$ the saturation intensity \cite{Met07}.
	The fraction $\frac{I_\text{laser}}{I_\text{sat}}$ is a free parameter and is the simulation equivalent to the laser intensity in the experiment.
	If an absorption occurs, the ion's velocity is reduced by $\hbar k_\text{L}/m_{\text{Be}}$ and it  enters the excited internal electronic state. 
	In the excited state the ion cannot absorb photons but can spontaneously decay to the electronic ground state with probability $\Gamma \Delta t$. Upon spontaneous decay, the ion receives an axial  velocity kick, $\hbar k/m_{\text{Be}} \cos(\varphi)$, where $\varphi$ is randomly distributed from 0 to 2$\pi$ to account for the randomness in the direction of the spontaneous emission. We note that this is a simplified model as it does not account for the anisotropy of spontaneous emission in a magnetic field \cite{Ita82}, which is however negligible for the studies in this paper.  Additionally, the ion can  emit photons by stimulated emission with probability $B_{21} (\dot{z}_\text{Be} ) \Delta t = B_{12}(\dot{z}_\text{Be} ) \Delta t$ where the ion's velocity is increased by $+\hbar k_\text{L}/m_{\text{Be}}$. 
	
	The above considerations enable us to accurately simulate the dynamics of the trapped particles and resonator including laser cooling of the beryllium ions under ideal conditions. However, in practice, several experimental uncertainties can occur. While long-term axial frequency drifts can easily be compensated, drifts on the order of the duration of one axial frequency measurement, typically about 60\,s, lead to frequency fluctuations that affect the energy exchange between the particles. The sources of these fluctuations are various, for example trapping voltage instabilities and drifts of the power supplies,  finite pressure and temperature  stabilities of the liquid helium bath, drifts due to the Coulomb-interaction between the   \Be{} ions, or residual inhomogeneities of the electric potential or magnetic field. A realistic estimate for the axial frequency fluctuation of the proton is $\sigma(\nu_z) \approx 50-100$\,mHz or $\sigma(\nu_z)/\nu_z \approx \SI{1e-7}{}$ at 60\,s for our experiment \cite{SchTh}.  
	In the following, we summarize the individual contributions by introducing an effective dynamical axial frequency variation by adding a modification $\Delta U$ to the trapping potential, where $\Delta U$  performs a random walk:
	\begin{equation}
		\Delta U_{n+1} = \Delta U_\text{n} + \delta U \cdot \mathcal{\tilde N}_\text{n}(0,1).
		\label{eq:voltage_stability1}
	\end{equation}
	The index $n$ indicates the step number and $\mathcal{\tilde N}_\text{n}(0,1)$ is another number drawn from a  Gaussian distribution. $\delta U$ is  a scaling factor determining the expectation value of the standard deviation of the random walk $\sigma(U_0)$ and is related to the frequency stability by $\sigma(U_0)/U_0 = 2 \sigma(\nu_z)/\nu_z$. 
	In order to express $\delta U$ in terms of practical experimental quantities, we define $\xi_\text{stab}$ as the relative voltage stability for frequency measurements of length 60\,s and obtain
	\begin{equation}
		%		\delta U = stability_voltage_p * U0_p * sqrt_steps_for_60s_inverse * gauss_distribution(rand_gen);
		\delta U_0 = \xi_\text{stab} \cdot U_0 \cdot \sqrt{\frac{1}{n(t=\SI{60}{s})}}.
		\label{eq:voltage_stability2}
	\end{equation}
	Here $n(t=\SI{60}{s})$ indicates the number of steps required for 60\,s of simulated time.  It should be noted that $\Delta U_\text{n}$ does not necessarily correspond to the actual behavior of the power supplies used in the experiment but is rather only a convenient tool to implement dynamical frequency fluctuations within one simulation run. 
	\label{sec:def_delta_unc}

	Data preparation and analysis is performed in \texttt{R} \cite{Rbasic}. Since typical simulated time spans are on the order of 30--60\,s, time steps of $\Delta t$ = 1--5\,ns imply about $10^{10}$ simulation steps, each with several dozens of operations. In order to minimize the computation time, the calculation-intensive part is outsourced to a compiled \texttt{C++} program  via the \texttt{Rcpp}-package \cite{Rcppcite}. In addition, the calculations are fully parallelized by the \texttt{doParallel}- and \texttt{foreach}-package \cite{foreach, doParallel}, allowing threads up to the number of virtual cores to run in parallel.
	
	In summary, we emphasize that the physics in our simulation is implemented from first principles. 
	We only use experimentally established model assumptions and no heuristic scaling factors.

	\section{Comparison of simulation results to experiment and theory}
	\label{sec:comparison}
\begin{figure*}
		\centering
		\centerline{
			\begin{tabular}[t]{ll}
				\rule{0pt}{-1.0ex}%
				\subf{\includegraphics[width=0.47\linewidth]{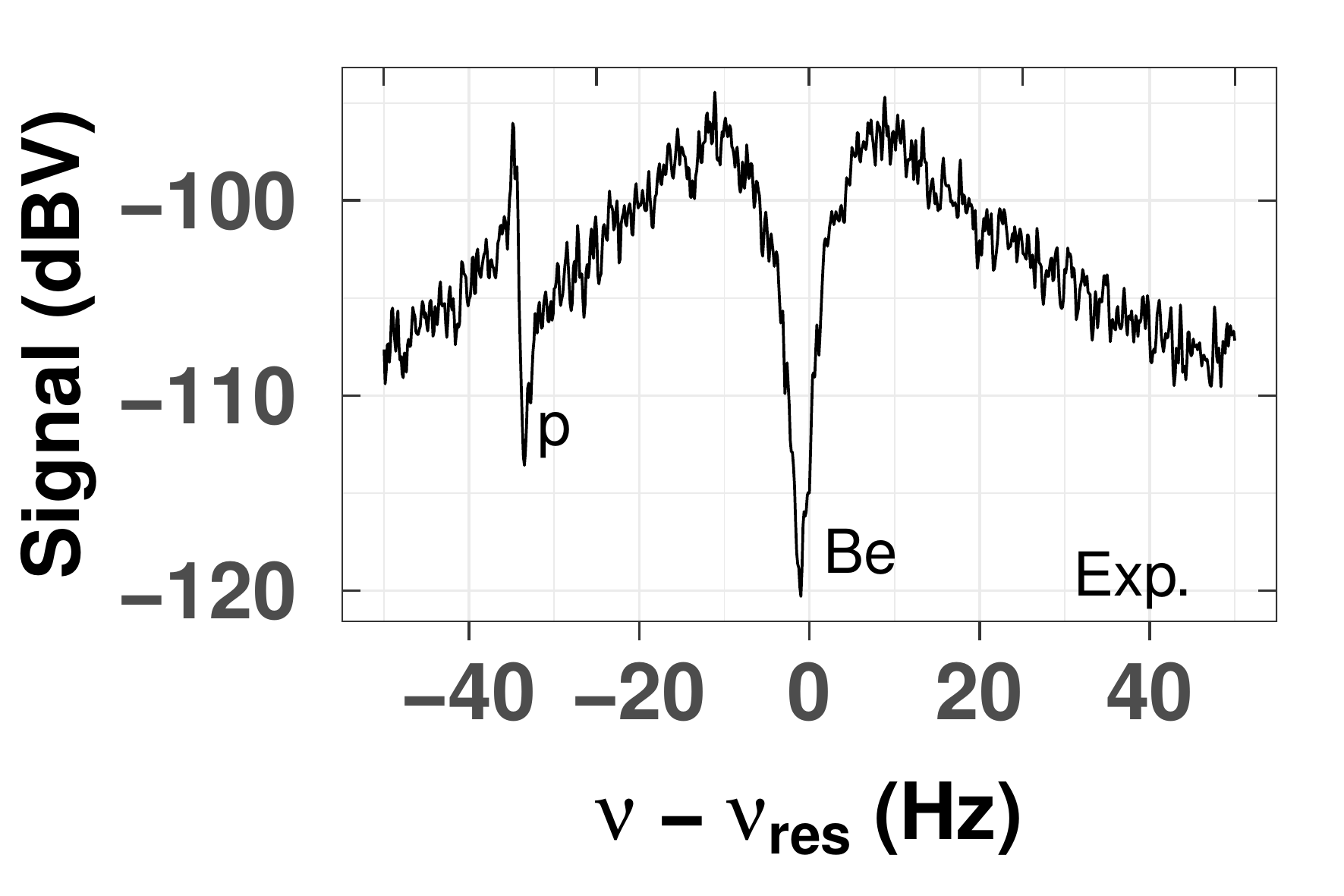}}{(a) }  & \subf{\includegraphics[width=0.47\linewidth]{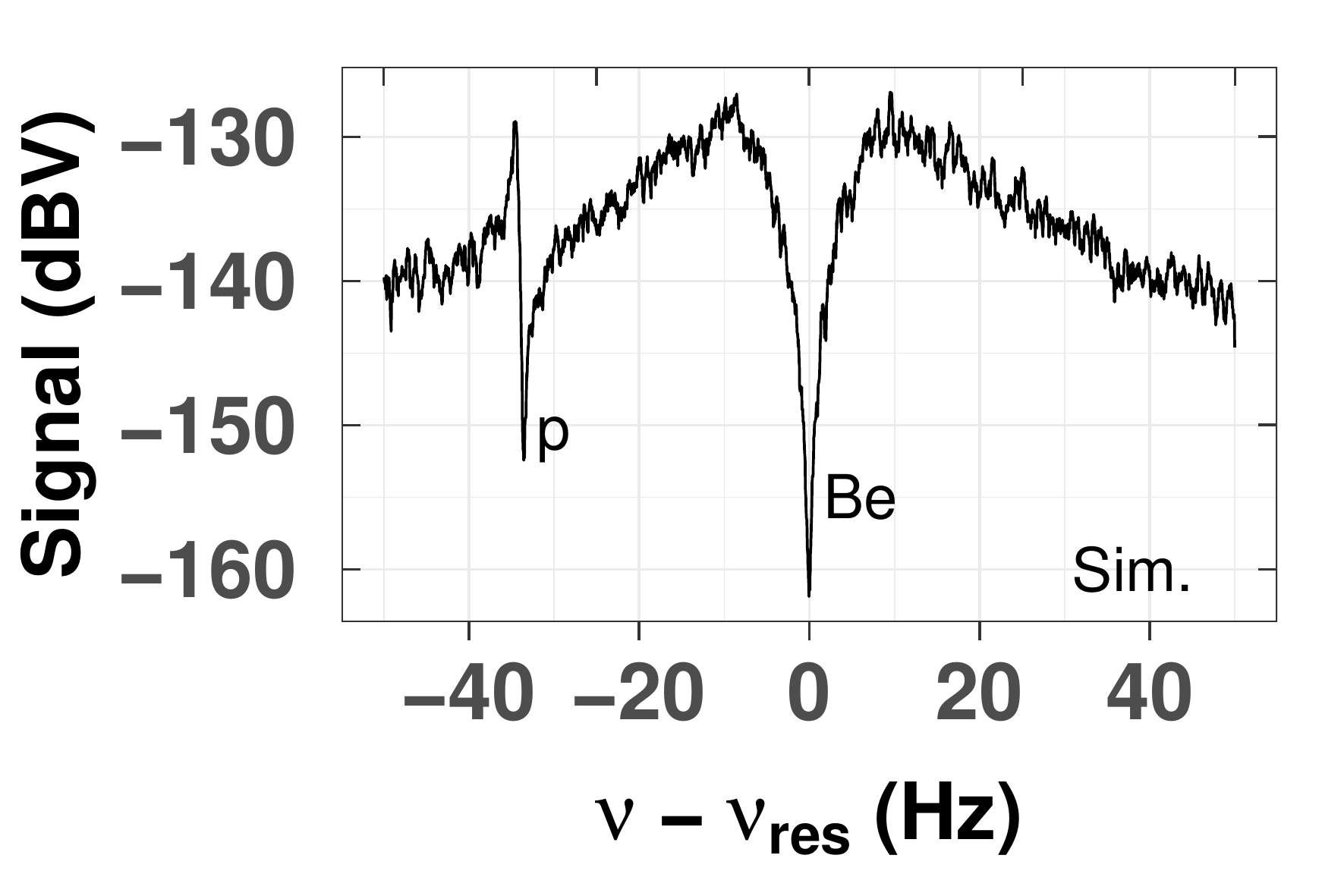}}{ (b) } \\
			\end{tabular}
		}
		\caption{ Example FFT spectra of the voltage drop across the RLC circuit. An experimental spectrum is shown in (a) and a simulated one with approximately matched parameters in (b). The broad peak is the resonance of the RLC circuit and the two dips  stem from a proton and several  \Be{} ions stored in independent traps, but  coupled to the same resonator. The proton is detuned from the detector by about -35\,Hz, leading to a dispersive dip. The dip widths are defined by the trap size, the number of ions in the trap and the ion masses \cite{Win75}.} 
		\label{fig:example_spectrum}
	\end{figure*}

	Next, we demonstrate that the simulation results agree well with theory and experiments, in particular those in Ref.~\cite{Boh21}. We simulate the \Be{} ions individually and consider no additional heating effects besides the one stemming from the resonator \cite{Bor19}. 
	Fig.\,\ref{fig:example_spectrum} shows an  FFT spectrum recorded with a common resonator involving the detector resonance, a proton in the PT detuned by -35\,Hz from the detector's  resonance frequency $\nu_\text{res}$ and several  \Be{} ions with $\nu_\text{z,Be}=\nu_\text{res}$ in the ST. Compared to the simulated spectrum in Fig.\,\ref{fig:example_spectrum}(b), the absolute signal level of the experimental spectrum in Fig.\,\ref{fig:example_spectrum}(a) is boosted by several amplifier stages.
	The broad resonance with full-width at half maximum (FWHM) $\gamma_\text{res}$ can be matched to the unperturbed lineshape of the noise-charged parallel RLC-circuit and the two dips stem from the proton and \Be{} ions.
	From such a dip spectrum several properties can be derived. First, the dip position yields the axial eigenfrequency of the particle. Here we confirm that the simulations produce the dip at the correct frequency  at  $\nu_\text{z} = 1/(2\pi) \sqrt{2qC_2U_0/m}$. Second, the dip width $\gamma_\text{z}$, defined as full-width at half maximum (FWHM) of the dip signal, is a measure  of the resonator-to-ion coupling strength. Assuming $\gamma_\text{z} \ll \gamma_\text{res}$, it can be calculated as
	\begin{equation}
		\gamma_z = \frac{1}{2\pi} \frac{R}{m} \frac{q^2}{D_\text{res}^2} N,
		\label{eq:dipwidth}
	\end{equation}
	where $N$ is the number of particles with mass $m$ \cite{Win75}.
	The dip widths of the simulated dip spectra match with Eq.\,(\ref{eq:dipwidth}). Similarly, the  simulated noise resonance has the same FWHM as we expect from theory. 
	Next, since the resonator is a noisy thermal bath, the particle energies follow a Boltzmann distribution with the mean temperature defined by the resonator. We confirm that both the absolute resonator temperature, which is a free input parameter in our simulations, as well as the Boltzmann distributions of the single particle energies are correctly reproduced. Regarding laser cooling, we  reproduce the theoretical Doppler limit of $T_D = \frac{\hbar \Gamma}{2 k_B}\approx 0.5$\,mK as minimal temperature of the  \Be{} ions. Furthermore, the scattering rate of photons $\gamma_s$ for a single  \Be{} ion artificially held at a constant velocity matches the theoretical one \cite{Met07}:
	\begin{equation}
		\gamma_s = \frac{I_\text{laser}}{I_\text{sat}} \frac{\Gamma/2}{1+ \frac{I_\text{laser}}{I_\text{sat}} + 4 \left( \frac{2\pi \delta_\text{L} + k_\text{L} \dot{z}_\text{Be} }{\Gamma}\right)^2 }.
	\end{equation}

	While the previously mentioned properties must be met for each oscillator individually and are straightforward to check, the coupling between them, especially between the proton and the  \Be{} ions, is more subtle. In order to reach our goal of $\approx \SI{10}{mK}$ axial proton temperature, it is crucial to understand the coupling mechanism in detail.
	In the following, we will compare the experimental data of Ref.\,\cite{Boh21} with  FFT spectra of simulation data for three different temperature regions.  Additionally, since the simulations provide us access to the time domain data, we can  demonstrate that our experimental signature is truly caused by a temperature manipulation. 
	
	Ref.\,\cite{Boh21} employed the coupling scheme with a common RLC resonator.  First, the coupling between the proton and  \Be{} ions was demonstrated above  the resonator temperature by bringing the proton to resonance with  excited beryllium ions. Therefore, the first temperature scale  at which we compare simulations against experiment is at particle temperatures above the resonator temperature, depicted in Fig.\,\ref{fig:3x3_hot}. Here we simulate a single proton and a cloud of 10  \Be{} ions. The beryllium ions are excited by a parametric drive with frequency $2\nu_\text{z,Be}$, which is implemented as an additional force $F_\text{dr}(t) = F_\text{0,dr} \cos(4\pi \nu_\text{z,Be} t)$ in Eq.\,(\ref{eq:finalset1b}). 
	In the experiment, special care was taken to ensure that the proton  was not directly affected by the heating drive but only via the coupling to the  \Be{} ions.
	Since  the axial frequencies of both particles are matched, not only the  \Be{} ions but also the proton significantly gains energy, demonstrating resonator-mediated energy exchange between the particles. Since $\gamma_\text{z,p} < \gamma_\text{z,Be} < \gamma_\text{res}$, the proton reaches a temperature higher than the one of the resonator. This results in a sharp peak above the resonator signal in the FFT spectrum in both the experimental and simulation FFTs, see Fig.\,\ref{fig:3x3_hot}(a) and (b), respectively. The time domain data of the simulations, depicted in Fig.~\ref{fig:3x3_hot}(c), support this explanation. Here, the heating drive is switched on at   $t=10$\,s. Before that, the energies of all three oscillators perform a random walk through  a Boltzmann distribution with the same temperature. With the parametric drive switched on, both the proton and  \Be{} ions gain energy equivalent to several hundred Kelvin. The average energy per ion is plotted for the beryllium cloud.

	Second, the coupling at the equilibrium temperature of the resonator was demonstrated.
	To this end, the characteristic response of  the coupled three-oscillator system is recorded, shown in Fig.\,\ref{fig:3x3_equil}. We employ a single proton and a single  \Be{} ion, whose frequencies are both detuned 130\,Hz from the resonator eigenfrequency. We then sweep the beryllium frequency over the proton frequency and display the signal power of the corresponding frequency spectra as colour-coded columns in a heat map. Both the experimental and simulated data in Fig.\,\ref{fig:3x3_equil}(a) and (b) exhibit the characteristic response of a coupled three-oscillator system.
	In addition, we plot the energy exchange between the proton, the  \Be{} ion and the resonator at 130\,Hz detuning in  the time domain for the simulated data in Fig.\,\ref{fig:3x3_equil}(c). To  emphasize the coupling between the proton and \Be{} ions, noise and dissipation were turned off by setting $R = \infty$ and $T_\text{res} = 0$ for this plot. Since the resonator FWHM $\gamma_\text{res}$ is $\approx \SI{30}{Hz}$, the resonator is only weakly excited by the particles. The exact behavior of the energy exchange in Fig.\,\ref{fig:3x3_equil}(c) strongly depends on the initial relative particle phase.
	
	Last, we establish common-resonator coupling  between a single proton and 50  laser-cooled  \Be{} ions. The frequency domain data in both experiment and simulation, Fig.\,\ref{fig:3x3_cold}(a) and Fig.~\ref{fig:3x3_cold}(b) respectively, show a broad dip with reduced signal-to-noise and a superimposed narrow dip below. The broad dip corresponds to the beryllium cloud which, due to the laser cooling, cannot fully compensate the resonator noise current  and hence exhibits a reduced signal strength. 
	The narrow dip corresponds to the proton,  which is coupled to both the resonator and the \Be{} cloud.
	In the time-domain plot,  Fig.\,\ref{fig:3x3_cold}(c), both particles have an initial  energy of 10\,\kkb. After turning on the laser at $t=7.5$\,s, not only the beryllium ions but also the proton is significantly cooled within a few seconds. This is exactly the effect we desire and this specific cooling scheme will be more thoroughly investigated in section \ref{sec:resonant_coupling}.

	 In this section, we have shown that our simulations are able to unambiguously reproduce the observed experimental signatures \cite{Boh21}. By giving access to the time domain data, the results provide another layer of understanding. 
	 The laser cooling simulations match the theoretical expectations down to the Doppler limit, and are therefore proven to be a practical and credible tool to investigate cooling schemes that achieve particle temperatures in the mK-regime.
	
	\setlength{\tabcolsep}{0pt}

	%	2x 4x6 inches
	\setlength\tabcolsep{-2.0pt}
	\newcommand{\myheight}{3.6cm}
	\newcommand{\addpicA}{\includegraphics[width=0.28\linewidth, height=\myheight]{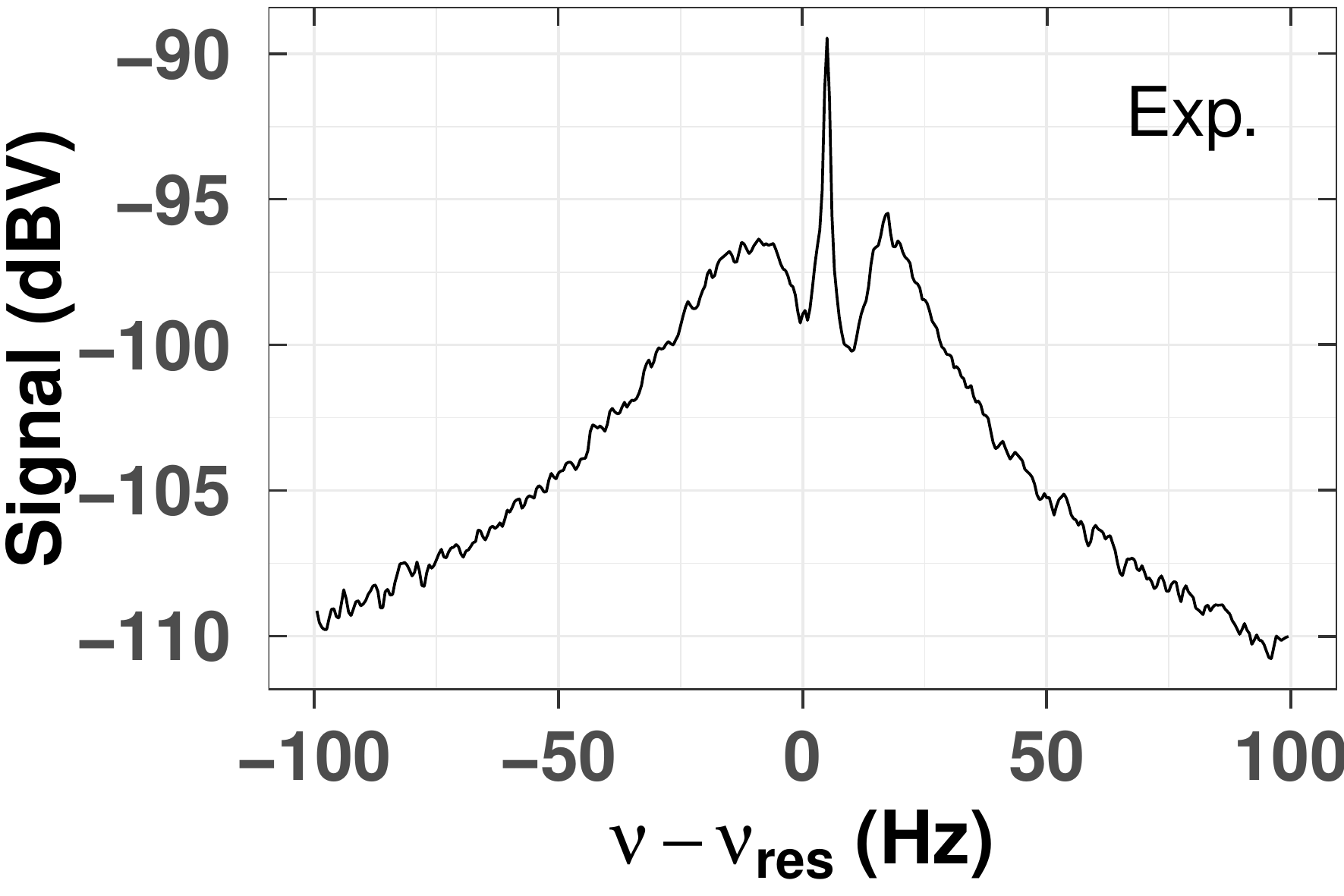}}
	\newcommand{\addpicB}{\includegraphics[width=0.28\linewidth, height=\myheight]{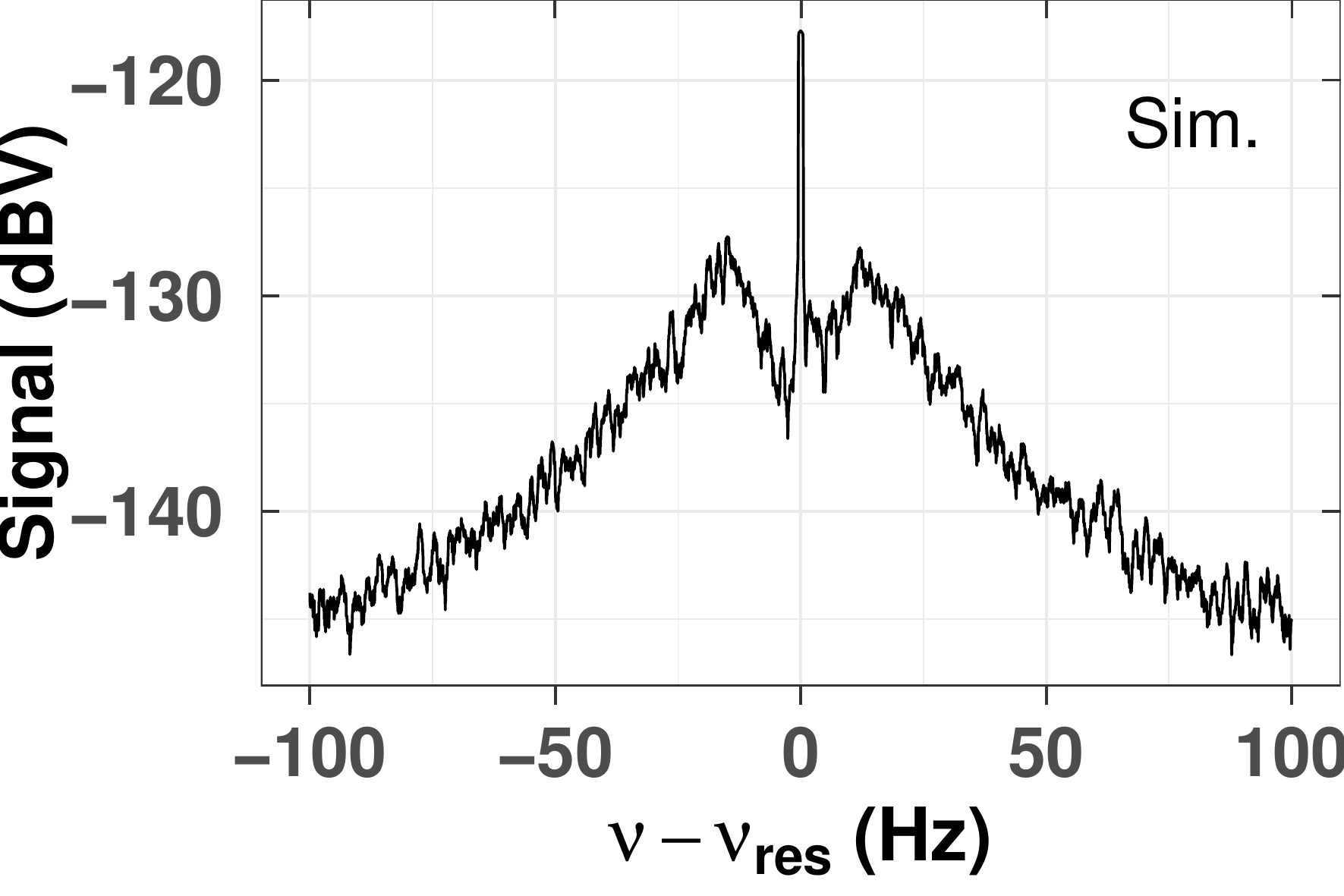}}
	\newcommand{\addpicC}{\includegraphics[width=0.28\linewidth, height=\myheight]{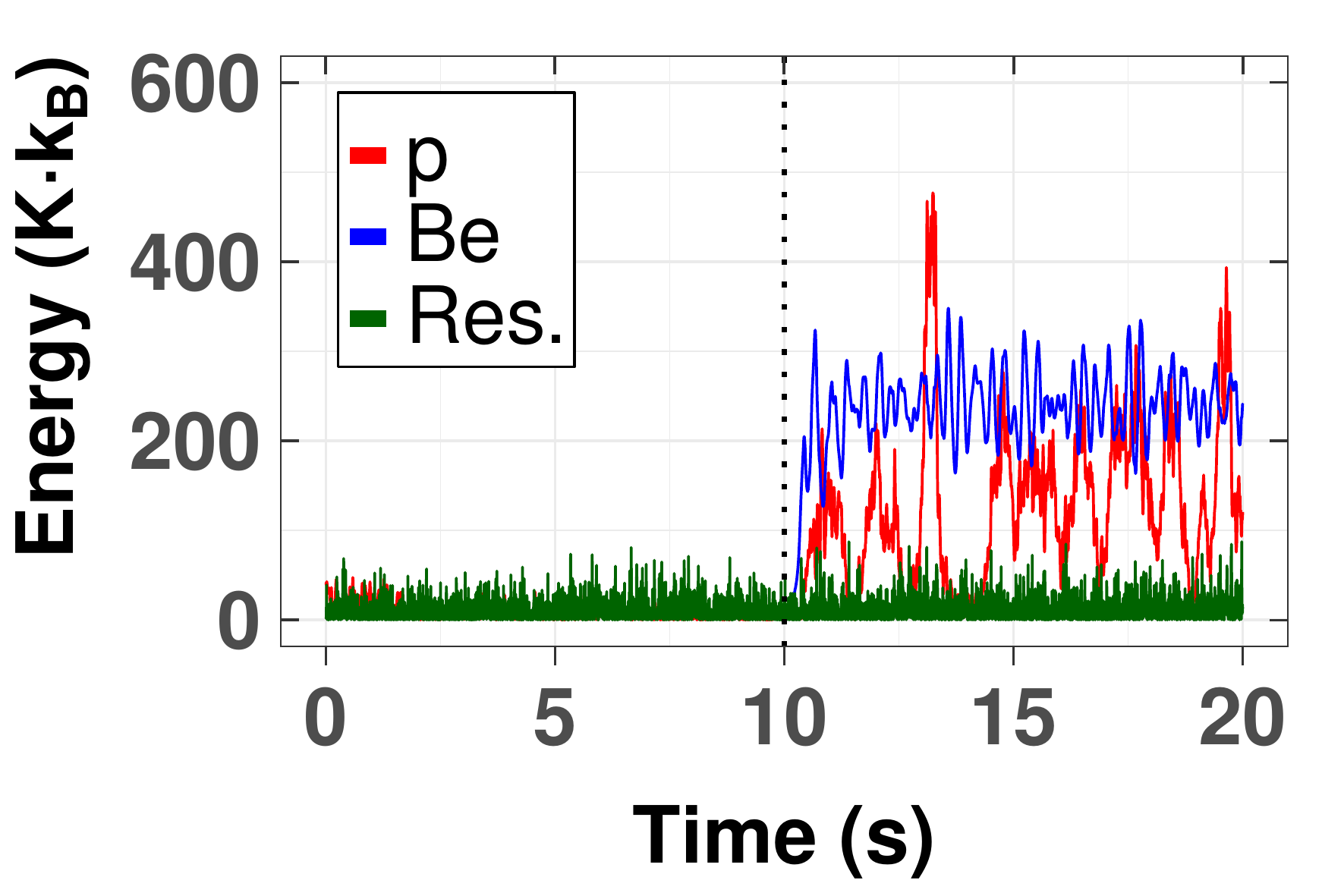}}
	
	%	4x8 und 4x6 inches
	\newcommand{\addpicD}{\includegraphics[width=0.30\linewidth, height=\myheight]{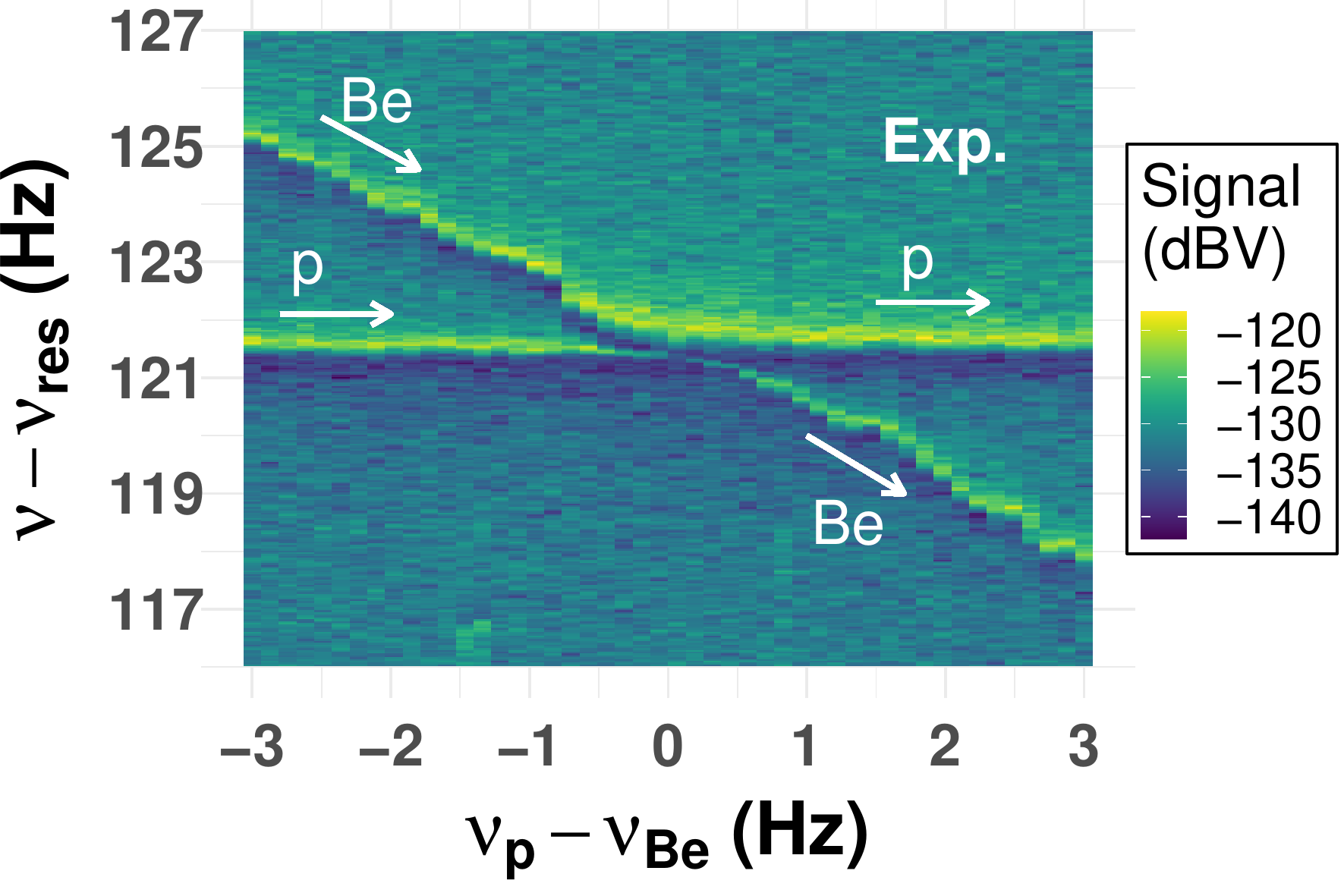}}
	\newcommand{\addpicE}{\includegraphics[width=0.30\linewidth, height=\myheight]{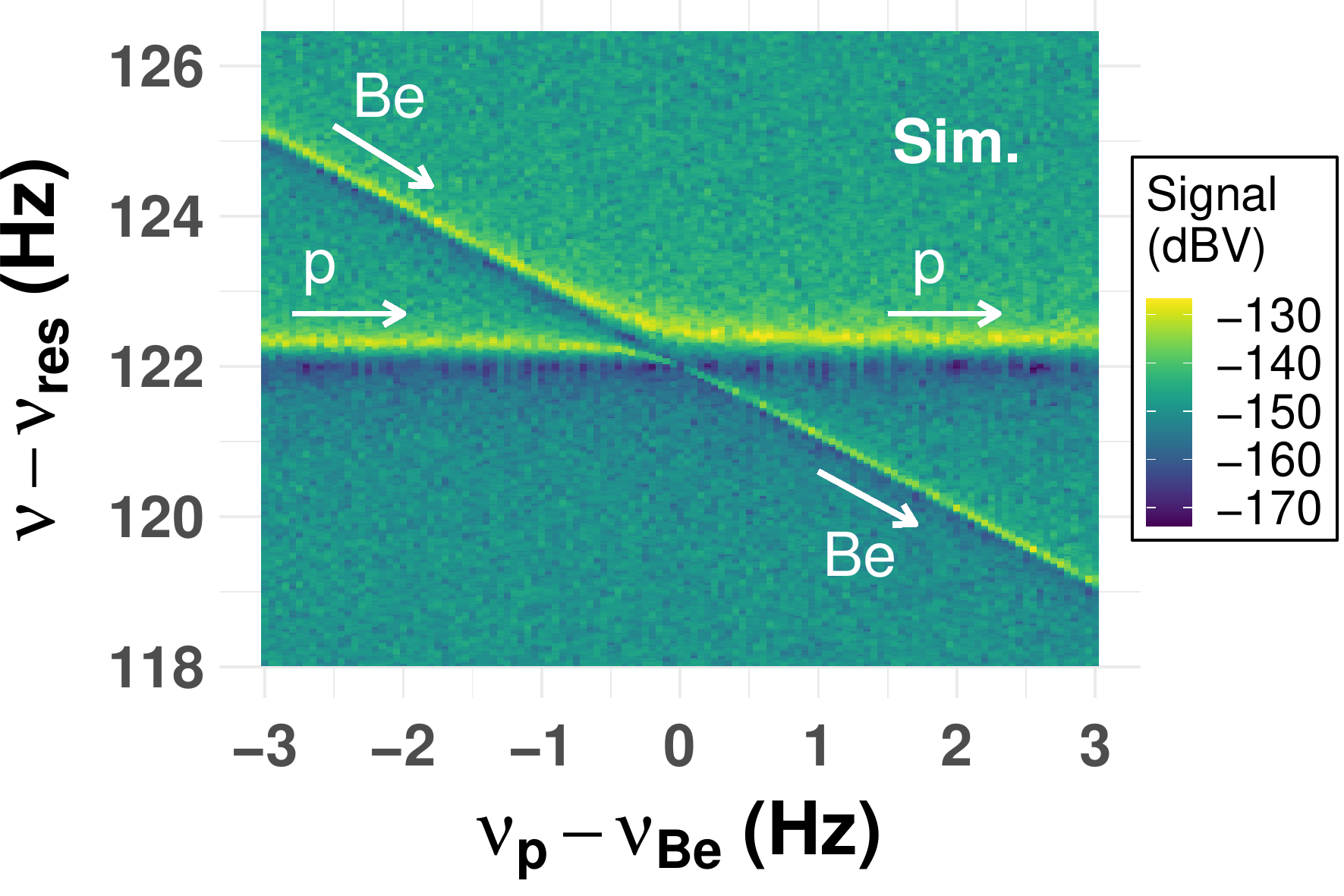}}
	\newcommand{\addpicF}{\includegraphics[width=0.28\linewidth, height=\myheight]{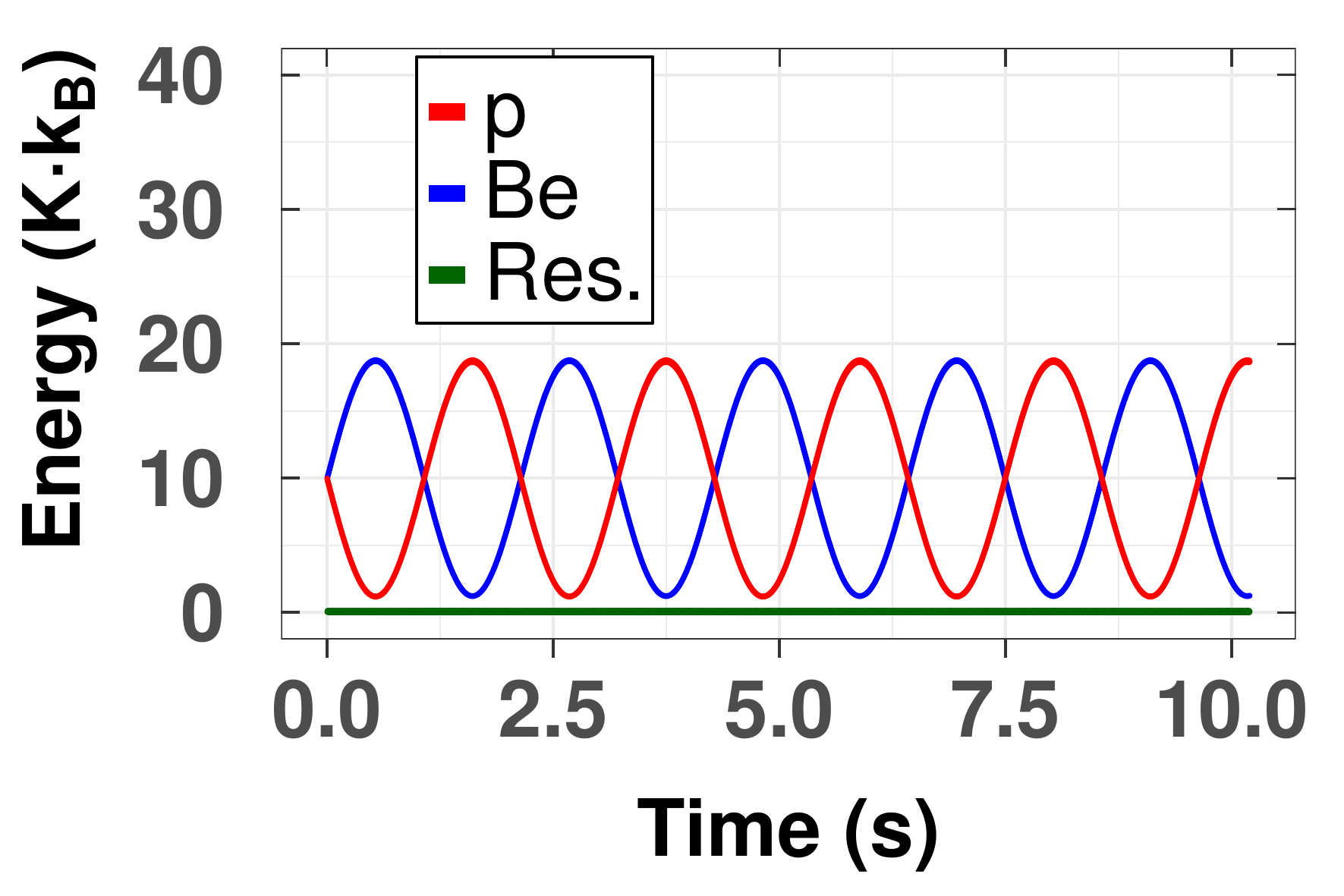}}
	
	\newcommand{\addpicG}{\includegraphics[width=0.28\linewidth, height=\myheight]{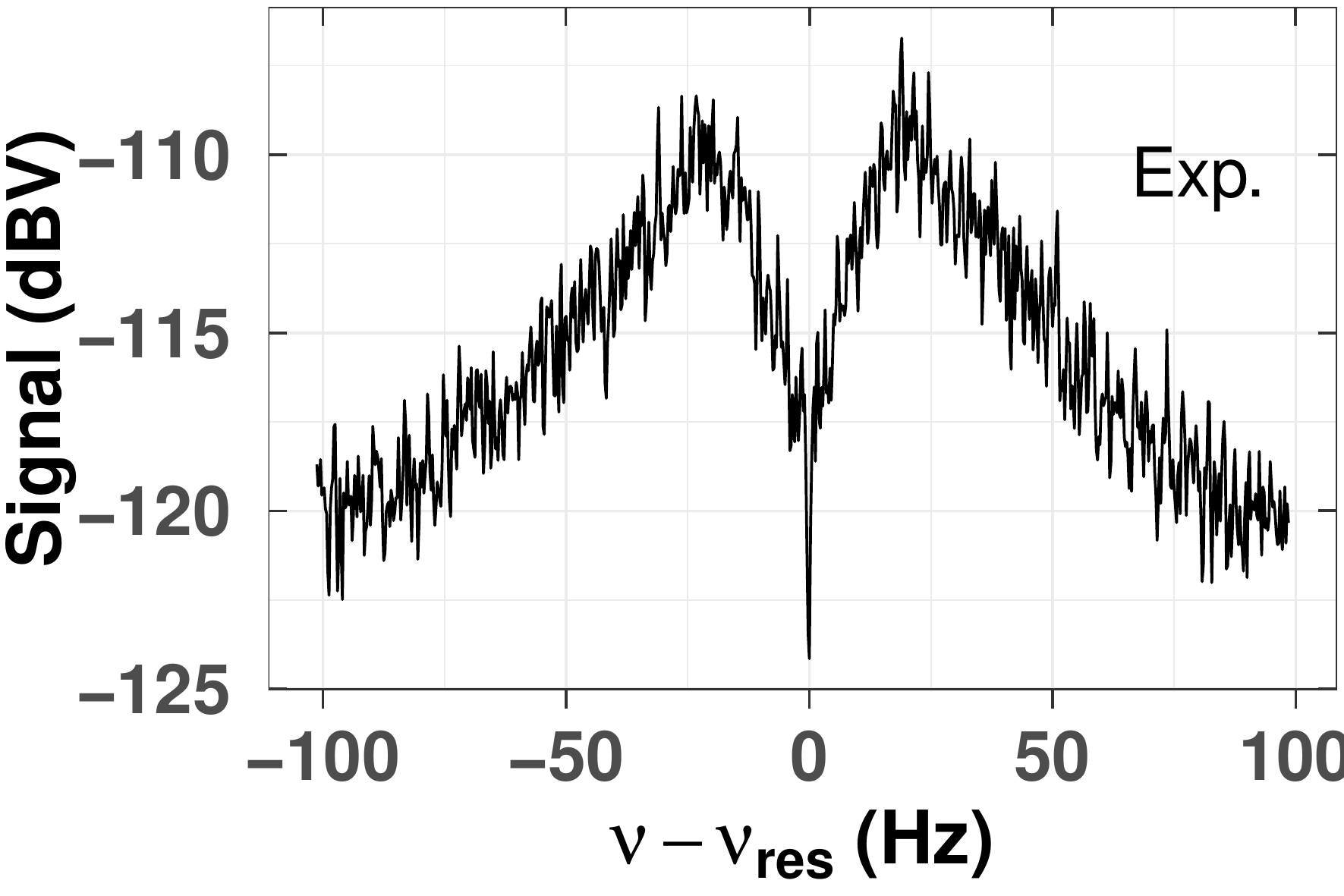}}
	\newcommand{\addpicH}{\includegraphics[width=0.28\linewidth, height=\myheight]{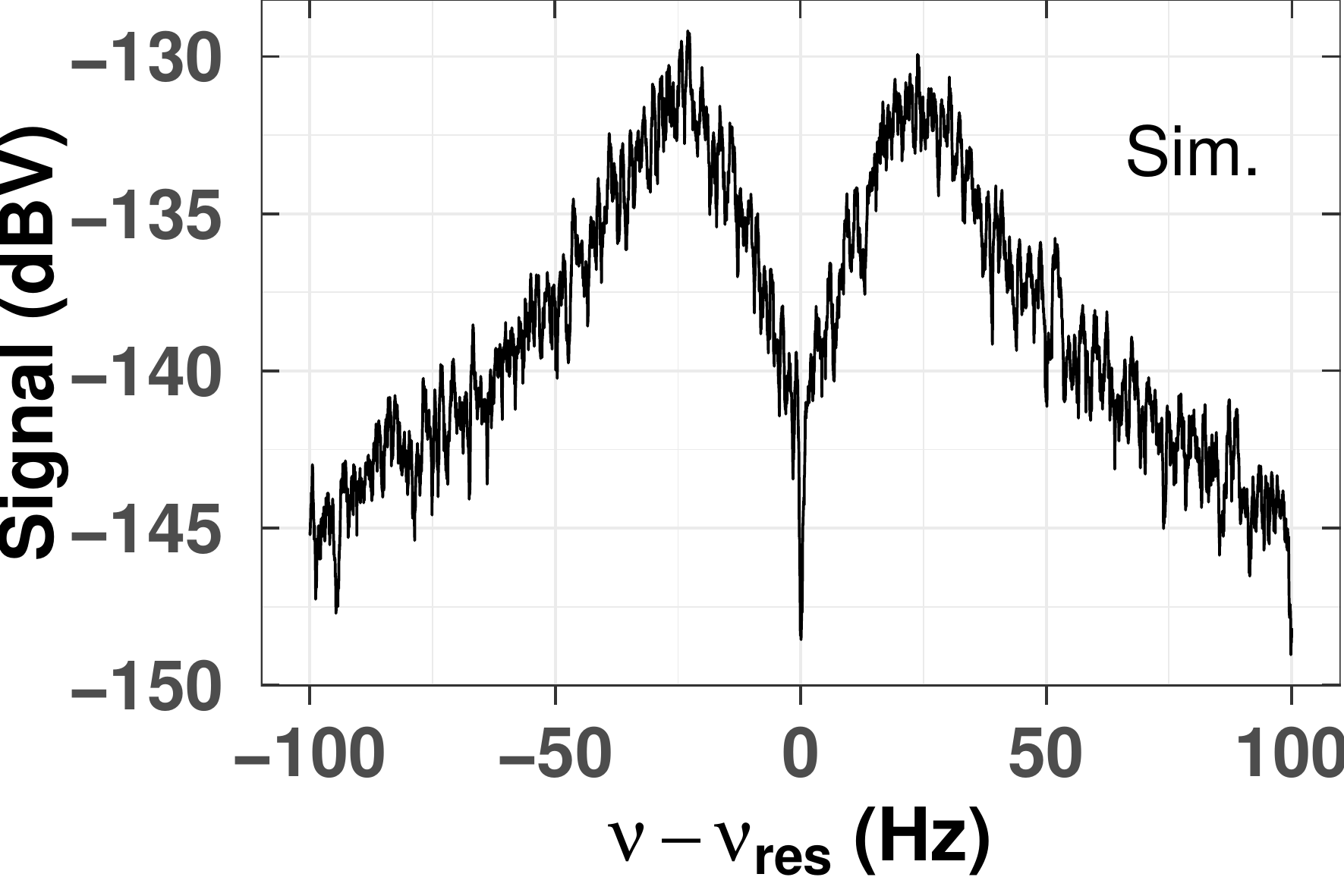}}
	\newcommand{\addpicI}{\includegraphics[width=0.28\linewidth, height=\myheight]{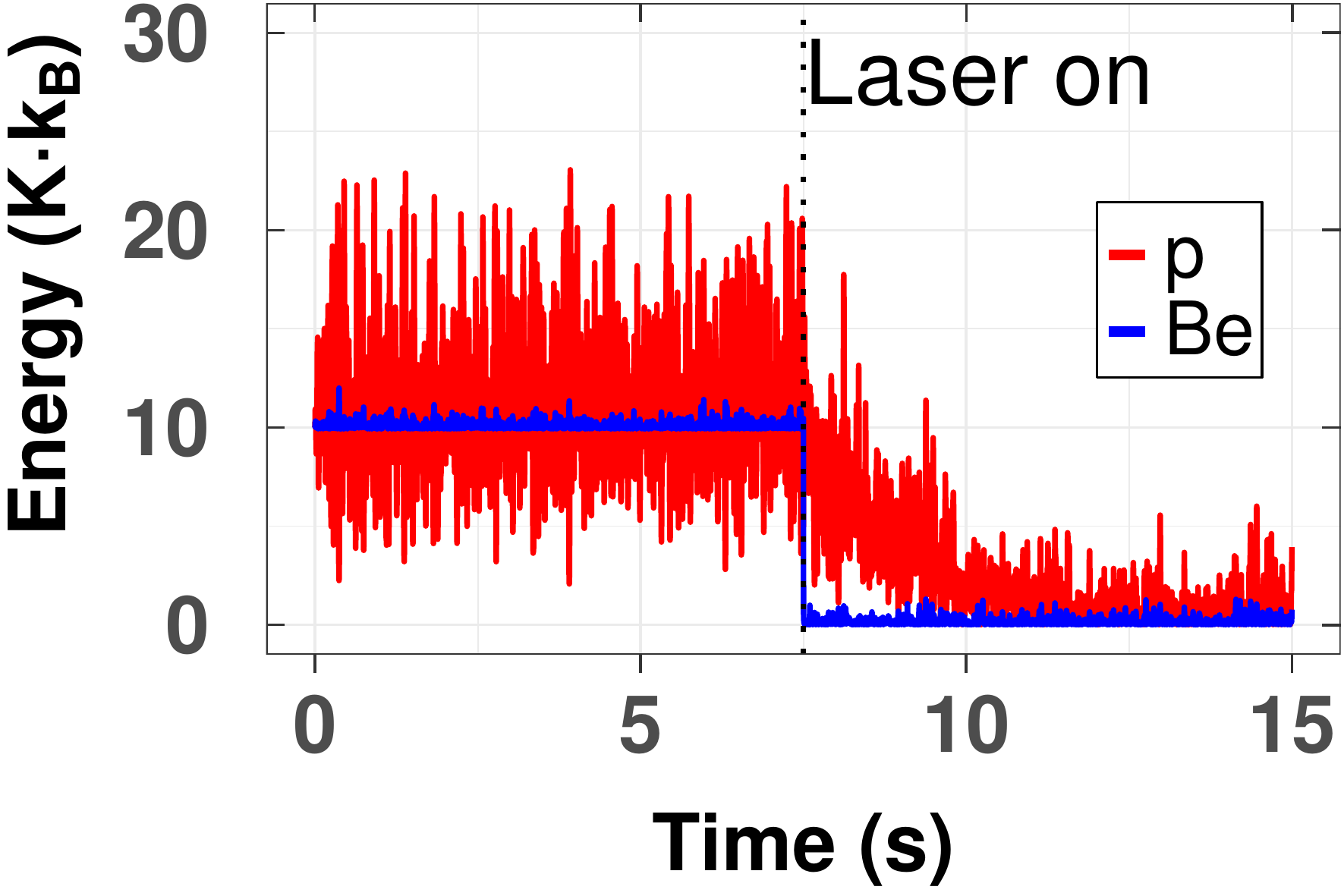}}

	\newgeometry{left=16mm, right=16mm}
	\begin{figure*}
		\centering
%		\centerline{
		
		\begin{tabular}[t]{lll}
			\rule{0pt}{-1.0ex}%
%			\centerline{\includegraphics[width=1.2\linewidth, height=\myheight]{01_excitation.pdf} }
%			\makebox[\textwidth][c]{\subf{\addpicA}{(a) }  & \subf{\addpicB}{ (b) } & \subf{\addpicC}{ (c) } }%
%			\makebox[\textwidth][c]{	\subf{\addpicA}{(a) }  & \subf{\addpicB}{ (b) } & \subf{\addpicC}{ (c) } }%
			\subf{\addpicA}{(a) }  & \subf{\addpicB}{ (b) } & \subf{\addpicC}{ (c) }
		\end{tabular}
%		}
		\caption{Comparison between experiment and simulation for $T_\text{p} \gg T_\text{res}$. The  \Be{} cloud is heated by a parametric $2\nu_z$-drive. 
			The proton is located in a different trap and is heated only due to the coupling to the \Be{} ions, creating the narrow peak signal in the FFT spectrum. (a) and (b) show the FFT spectrum of the experiment and simulation, respectively. The simulated effect of switching on the heating drive at $t=\SI{10}{s}$ (dashed line) is demonstrated in (c).  The energy per ion is depicted for the beryllium cloud.   }
		\label{fig:3x3_hot}
	\end{figure*}

	\begin{figure*}
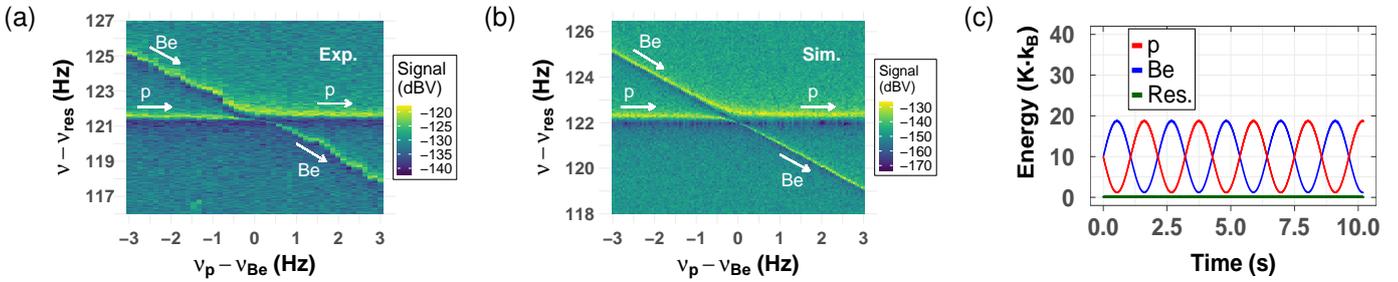

		\centering
		\begin{tabular}[t]{lll}
			\rule{0pt}{-1.0ex}%
			\subf{\addpicD}{(a) } & \subf{\addpicE}{ (b) } & \subf{\addpicF}{ (c) } 
		\end{tabular}
		\caption{The coupling between the proton and \Be{} is demonstrated at $T_\text{p} = T_\text{res}$ by sweeping $\nu_\text{z,Be}$ over $\nu_\text{z,p}$. The experiment data in (a) and simulation data in (b) show the characteristic response of a coupled three-oscillator system. In (c) we show simulated time domain data for the simplified case with no resonator noise.}
		\label{fig:3x3_equil}
	\end{figure*}

	\begin{figure*}
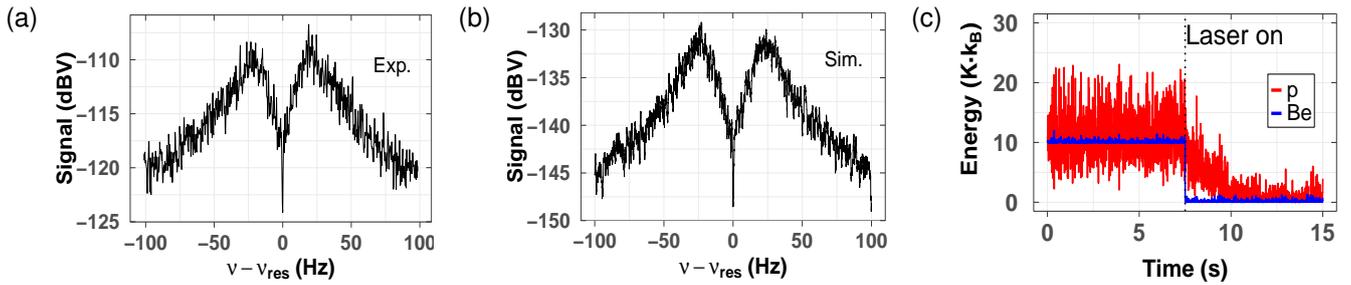

		\centering
		\begin{tabular}[t]{lll}
			\subf{\addpicG}{(a) } & \subf{\addpicH}{ (b) } & \subf{\addpicI}{ (c) } 
		\end{tabular}
		\caption{Comparison between experiment and simulation for $T_\text{p} < T_\text{res}$. Both the experiment data in (a) and the simulation data in (b) show  a common-resonator FFT with  a  laser-cooled beryllium cloud (broad dip with reduced signal strength) superimposed  by a single proton (narrow dip). In (c) the cooling laser is switched on at $t=\SI{7.5}{s}$, making the cooling effect on the proton visible.  The energy per ion is depicted for the beryllium cloud.   }
		\label{fig:3x3_cold}
	\end{figure*}
\restoregeometry

	\section{Coupling scenarios}
	\label{sec:coupling_scenarios}
	
	Next, we will theoretically investigate the  three coupling scenarios whose experimental setups were described in Sec.\,\ref{sec:coupling_mechanisms}, namely common-endcap coupling, resonant coupling via a common RLC circuit, and off-resonant coupling via a common RLC circuit.
	
	We simulate individual beryllium ions in Sec.~\ref{sec:resonant_coupling} and effective clouds otherwise since only in Sec.~\ref{sec:resonant_coupling} absolute laser intensities are of interest. The initialization of the particle modes is randomized and has no influence on the coupling schemes.  
	If not explicitly stated otherwise, we use the following simulation parameters for the resonator: $R=\SI{236e6}{\Omega}$, $L=\SI{3}{mH}$, $C_\text{res} = \SI{21.4}{pF}$, yielding a resonator Q-factor of \SI{20000} and a  resonance frequency of \SI{626.9}{kHz}. Regarding the trap parameters, we adjust the product $C_2 U_0$ such that the particle oscillates at the desired frequency. The effective electrode distances are $D_\text{res,p} = \SI{10.0}{mm}$ for the PT, $D_\text{res,Be} = \SI{5.7}{mm}$ for the ST and $D_{\text{com,p}} = D_\text{com,Be} = \SI{3.2}{mm} $ for the CT and BT 
	(compare Fig.~\ref{fig:experimental_setup_together}) \cite{SchTh}. 
	The common endcap capacitance is $C_\text{com} = \SI{5.5}{pF}$. Additionally, we assume that the resonator has an effective noise temperature of 10\,K, accounting for the fundamental Johnson-Nyquist noise at 4.2\,K (the physical temperature of the experiment) and the input noise of our cryogenic amplifiers \cite{Nag16, Ulm09}. 
	Finally, a time step of  $\Delta t = \SI{4}{ns}$ was used for all subsequent simulations.

	\subsection{Common-endcap coupling}
	\label{sec:capacitive_coupling}

	In order to  simulate the common-endcap coupling in our trap system, we detune the resonator eigenfrequency by $200$\,kHz so that the energy exchange is dominated by the common-endcap coupling. In the experiment, this will be achieved by adding or subtracting an additional parallel capacitance to the resonator using a cryogenic switch \cite{WieTh}. This has the advantage that the particle frequencies can be matched on resonance and no adjustment of the trapping potentials is necessary when starting the coupling procedure.
	Since the resonator terms in Eq.\,(\ref{eq:finalset1a}) - (\ref{eq:finalset1d}) are then negligible for the ion-ion-coupling with no laser applied, the system of coupled differential equations can be solved  analytically. 
	In order to derive a general solution we allow for more than one particle in each cloud by setting $q_\text{p} \rightarrow N_\text{p}\, q_\text{p}$, $m_\text{p} \rightarrow N_\text{p}\, m_\text{p}$,  $q_{\text{Be}} \rightarrow N_{\text{Be}} \, q_\text{Be}$ and $m_{\text{Be}} \rightarrow N_{\text{Be}} \, m_{\text{Be}}$ in Eq.\,(\ref{eq:finalset1c},  \ref{eq:finalset1d}), where $N_\text{p,Be}$ is the number of protons and beryllium ions, respectively. For the sake of a compact notation we abbreviate $D_\text{com,p} \equiv D_\text{p}$ and $D_\text{com,Be} \equiv D_{\text{Be}}$ in the following calculations. 	Furthermore, we assume that the particle eigenfrequencies are exactly matched, i.e. $\frac{2q_\text{p} C_\text{2,p} U_\text{0,p}}{m_\text{p}} = \frac{2q_\text{Be} C_\text{2,Be} U_\text{0,Be}}{m_\text{Be}} = \omega_0^2$. Solving the coupled differential equation for the proton motion yields 
	\begin{align}
		z_\text{p}(t) = c_1 e^{i \lambda_1 t} +  c_2 e^{i \lambda_2 t} +  c_3 e^{i \lambda_3 t} +  c_4 e^{ i \lambda_4 t}
	\end{align}
 	where $c_{1,2,3,4}$ are free coefficients determined by the initial conditions and $\lambda_{1,2,3,4}$ are the corresponding eigenfrequencies:
 	\begin{align}
 		\lambda_{1,2} &= \pm \, \omega_0 \\
 		\lambda_{3,4} &= \pm \sqrt{ \omega_0^2 - \frac{ m_\text{p} D_\text{p}^2 N_{\text{Be}} q_\text{Be}^2 + m_{\text{Be}} D_{\text{Be}}^2 N_\text{p} q_\text{p}^2 }{C_{\text{com}} D_\text{p}^2 D_{\text{Be}}^2 m_\text{p} m_{\text{Be}} } }  \\
 		&\approx \pm (\omega_0 - \Omega ).
 		 		\label{eq:lambdas}
 	\end{align}
	Here we abbreviated
		\begin{equation}
		\Omega = \frac{1}{2} \frac{ m_\text{p} D_\text{p}^2 N_{\text{Be}} q_\text{Be}^2 + m_{\text{Be}} D_{\text{Be}}^2 N_\text{p} q_\text{p}^2 }{ \omega_0 C_{\text{com}} D_\text{p}^2 D_{\text{Be}}^2 m_\text{p} m_{\text{Be}} }
		.
		\label{eq:CC_rabi}
	\end{equation}
	
	Assuming the initial conditions $z_\text{p} = z_0$, $\dot z_\text{p} = 0$, $z_{\text{Be}} = 0$ and  $\dot z_{\text{Be}} = 0$, the temporal evolution of the energy of the proton is 
	\begin{align}
		\label{eq:CC_energy_exchange}
		E_\text{p}(t) =& \frac{1}{2}  \frac{m_\text{p} z_0^2 }{(m_\text{p} D_\text{p}^2 N_{\text{Be}} q_\text{Be}^2 + m_{\text{Be}}D_{\text{Be}}^2 N_\text{p} q_\text{p}^2)^2} \left( \vphantom{\frac{x}{y} }\right. \omega^2 \left[  (m_\text{p}  D_\text{p}^2 N_{\text{Be}} q_\text{Be}^2 )^2 + \right.  \\   
		 &    \left. \left. (m_{\text{Be}} D_{\text{Be}}^2 N_\text{p} q_\text{p}^2)^2 + 2m_\text{p} m_{\text{Be}} N_\text{p} N_{\text{Be}} q_\text{p}^2 q_\text{Be}^2D_\text{p}^2 D_{\text{Be}}^2 \cos(\Omega t) \right]  + \mathcal{O}(\Omega \omega) \vphantom{ \frac{x}{y} } \right). \nonumber
	\end{align}

	 	\begin{figure*}
		\centering
		\centerline{
			\begin{tabular}[t]{ll}
				\rule{0pt}{-1.0ex}%
				\subf{\includegraphics[width=0.47\linewidth]{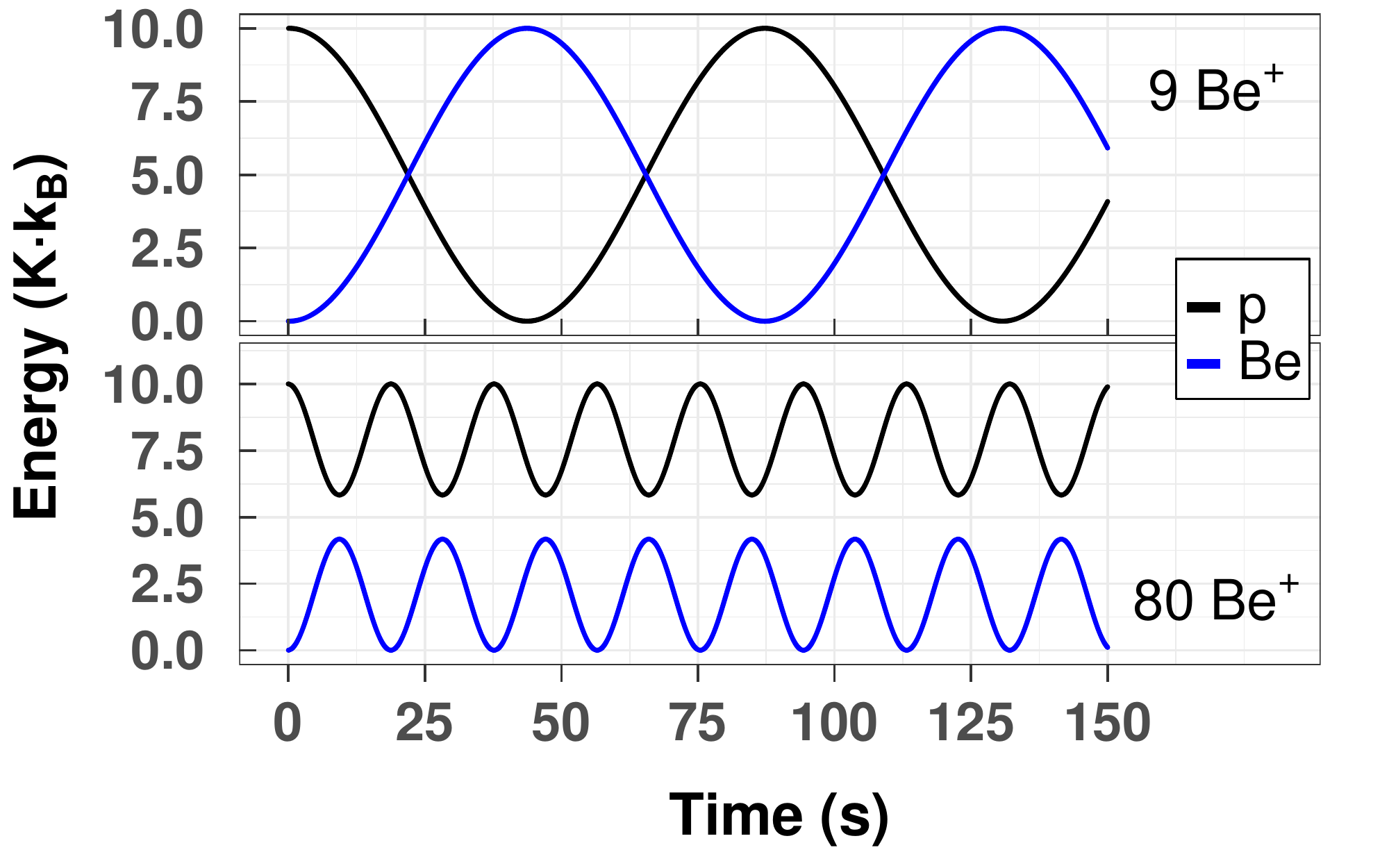}}{ (a) }  &
				\subf{\includegraphics[width=0.47\linewidth]{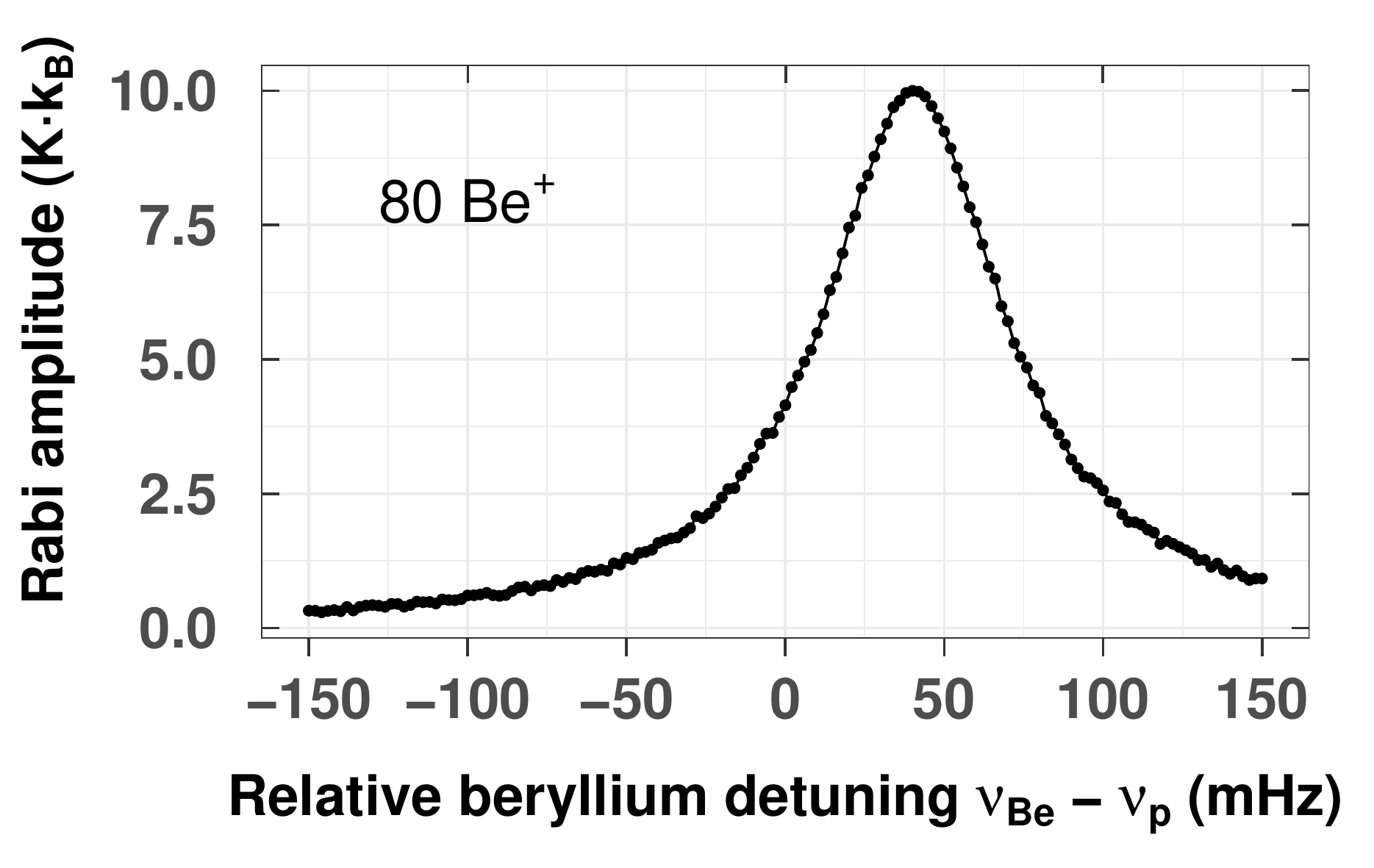}}{(b) } \\
%				\subf{\includegraphics[width=0.47\linewidth]{CC_energy_exchange_example}}{(b) } \\
			\end{tabular}
		}
		\caption{(a) Simulated energy of a single proton capacitively coupled to a cloud of  \Be{} ions. In the case of $D_{\text{com,p}} = D_{\text{com,Be}}$ the energy is exchanged completely only for 9 beryllium ions (upper plot). In contrast, for 80 beryllium ions the energy exchange is incomplete. (b) Energy exchange amplitude as a function of relative particle detuning for 80 \Be{} ions. The  maximum Rabi oscillation amplitude is restored by detuning the beryllium ions relative to the proton by  $+40$\,mHz.}   
		\label{fig:CC_energy_exchange}
	\end{figure*}

	 Since $\Omega \ll \omega$, the terms with $\mathcal{O}(\Omega \omega)$ can be neglegted. 
	 From Eq.\,(\ref{eq:CC_energy_exchange}) $\frac{\Omega}{2\pi}$ can be identified as the Rabi frequency of the coupled proton-beryllium system. The difference between Eq.~(\ref{eq:CC_rabi}) and previously reported versions \cite{Boh17, Tu21} arises from the fact that in this  work, up to Eq.\,(\ref{eq:lambdas}) no approximations were used. In order for the energy to be exchanged completely, i.e. $E_\text{p}(t=\pi/\Omega) = 0$, the condition 
	 \begin{equation}
		\frac{N_\text{p} q_\text{p}^2}{m_\text{p} D_\text{p}^2} \overset{!}{=} \frac{N_\text{Be} q_\text{Be}^2}{m_\text{Be} D_\text{Be}^2}
	 	\label{eq:CC_condition}
	 \end{equation}
 	must be satisfied. Inserting this condition into Eq.\,(\ref{eq:CC_rabi}) reproduces the previously reported formula  \cite{Boh17, Tu21} for the Rabi oscillation frequency. Eq.~(\ref{eq:CC_condition}) reflects the simultaneous conservation of energy and conservation of charge on the common-endcap capacitance within one Rabi cycle. We note that  the studied system of coupled oscillators is symmetric in many experiments and Eq.~(\ref{eq:CC_condition}) is automatically fulfilled. 
 	However, we emphasize that if the condition in Eq.\,(\ref{eq:CC_condition}) is not met, the energy is exchanged only partially and with a faster frequency. As demonstrated in the following, this has considerable implications for the realization of the cooling schemes in the experiment. 
 	Fig.\,\ref{fig:CC_energy_exchange}(a) shows an exemplary energy exchange between a single proton and 9 or 80  \Be{} ions, top and bottom plot, respectively. Since $D_\text{p} = D_\text{Be}$ and $m_\text{Be} \approx 9\,m_\text{p}$ in our case, the energy exchange is complete for 9 beryllium ions. While the Rabi frequency for 80 beryllium ions is higher, only a fraction of the proton's energy is exchanged. This would be a severe limitation of the  common-endcap coupling technique as one would need to choose between a complete energy exchange ($N_{\text{Be}} = 9$) or a fast coupling time ($N_{\text{Be}} \gg  9$). However, the behavior of the Rabi oscillations when varying the number of beryllium ions is equivalent to introducing a small relative frequency detuning between the two species. Consequently, one can reverse the effect of using $N_{\text{Be}} \gg 9$  by introducing a small relative detuning. The analytical calculations are  laborious, thus we make use of the numerical simulations here.
 	Fig.\,\ref{fig:CC_energy_exchange}(b) shows the Rabi oscillation amplitude (peak-to-peak) for 80  \Be{} ions as a function of relative particle detuning. While for an exact frequency matching only $\approx 4$\kkb{} are exchanged per Rabi cycle, at +40\,mHz  relative detuning the full Rabi amplitude is recovered.  In the case of nonzero detunings Eq.~(\ref{eq:CC_rabi}) does not hold anymore. However, if the detuning is chosen such that the energy is exchanged completely,  the previously reported $\Omega \propto \sqrt{N_\text{Be}}$ is preserved, i.e. $\Omega(\SI{40}{mHz}, 80\,\text{Be}) =  \sqrt{\frac{80}{9}} \Omega(\SI{0}{mHz}, 9\,\text{Be})$. 
 	The parameters presented here yield $\Omega(\SI{0}{mHz}, 80\,\text{Be})/(2\pi) = \SI{53}{mHz}$ and $\Omega(\SI{40}{mHz}, 80\,\text{Be})/(2\pi) = \SI{34}{mHz}$.
 	Although matching two species to a few  ten mHz is experimentally challenging, it is crucial to account for this in the experimental realization of the coupling scheme as it impacts the cooling times and the achieved temperature limits. 
% 	  would require to match and stabilize the particle frequencies to a level fo 10\,mHz 
 	
 	\begin{figure*}
 	\centering
 	\centerline{
 		\begin{tabular}[b]{ll}
 			\rule{0pt}{-1.0ex}%
 			\subf{\includegraphics[width=0.47\linewidth]{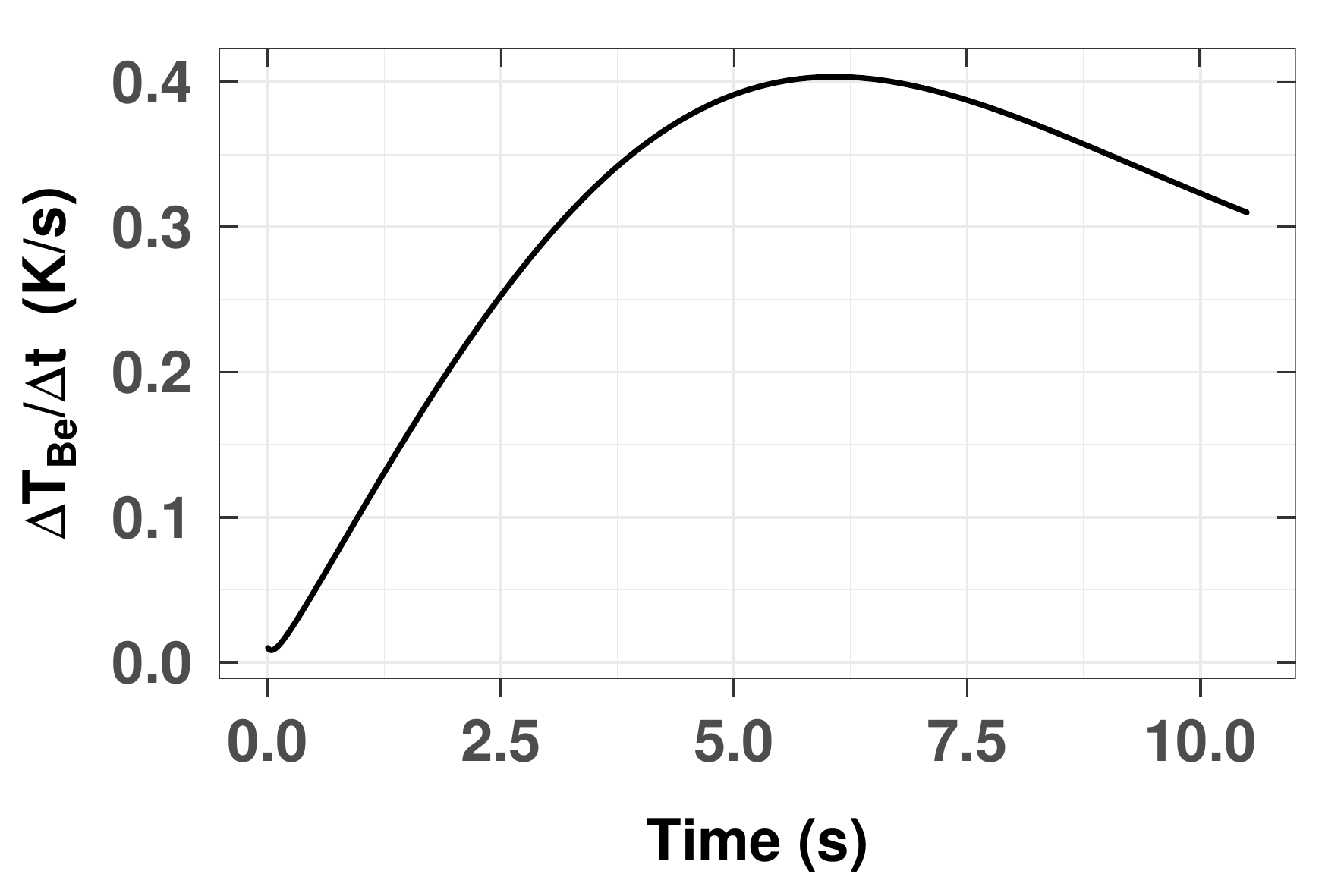}}{ (a) }  & 
 			\subf{\includegraphics[width=0.47\linewidth]{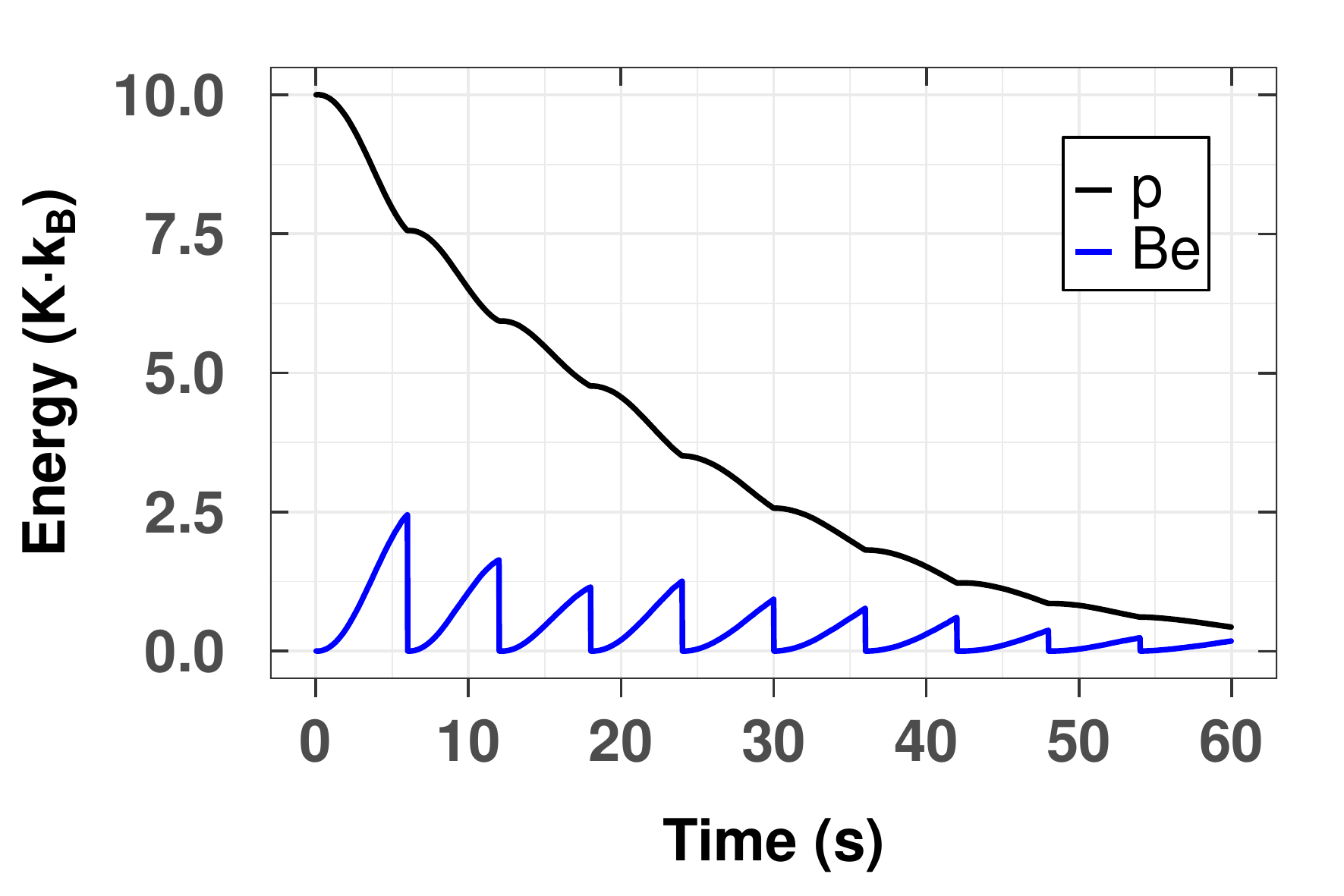}}{ (b) } \\
 		\end{tabular}
 	}
 	\caption{(a) Evolution of the simulated energy gain of the  \Be{} cloud divided by the time after turning off the laser for the common-endcap coupling.
 		The depicted data is an average temperature for initial axial frequency mismatches of $\Delta_\text{unc} = \SI{100}{mHz}$ and a dynamical stability of $\xi_\text{stab} = \SI{1e-7}{}$ as defined in Sec.~\ref{sec:def_delta_unc}. The maximum energy exchange rate is reached at  6.0\,s for 80 Be$^+$ ions, defining the optimal cycle time for the first laser pulse. 
 		(b) Representative example of the cooling process with a laser pulse applied every  6.0\,s.     }
 	\label{fig:CC_laser_duty}
 \end{figure*}
 
 	The initial proposals \cite{Boh17, Hei90} envisaged  cooling close to the Doppler limit by coupling  the target ion for exactly half a Rabi cycle (one $\pi$-pulse) to the  subsequently laser-cooled $^9$Be$^+$-cloud.
 		However, Fig.~\ref{fig:CC_energy_exchange}(b)  shows  that a frequency mismatch of only $\approx 10$\,mHz  limits the residual proton energy after one $\pi$-evolution to $\gtrsim 1$\kkb, far higher than the 10\,mK required for our experiment. 
 	 As a result,  we investigate a cooling scheme that is based on repetitively turning    on the cooling laser  and performing several imperfect energy exchanges.
 	 Compared to continuous laser operation,   chopping the laser achieves orders of magnitude faster cooling. The reason for this can be understood in the Rabi oscillation picture: In the continuous case the  \Be{} cloud is artificially held at its minimal temperature. The cooling rate for the proton is given by the local slope of the energy exchange curve,  i.e. it vanishes to first order for continuously cooled \Be{} ions \cite{Tu21}.

\begin{figure*}
	\centering
	\centerline{
		\begin{tabular}[tb]{ll}
			\rule{0pt}{-1.0ex}%
			\subf{\includegraphics[width=0.47\linewidth]{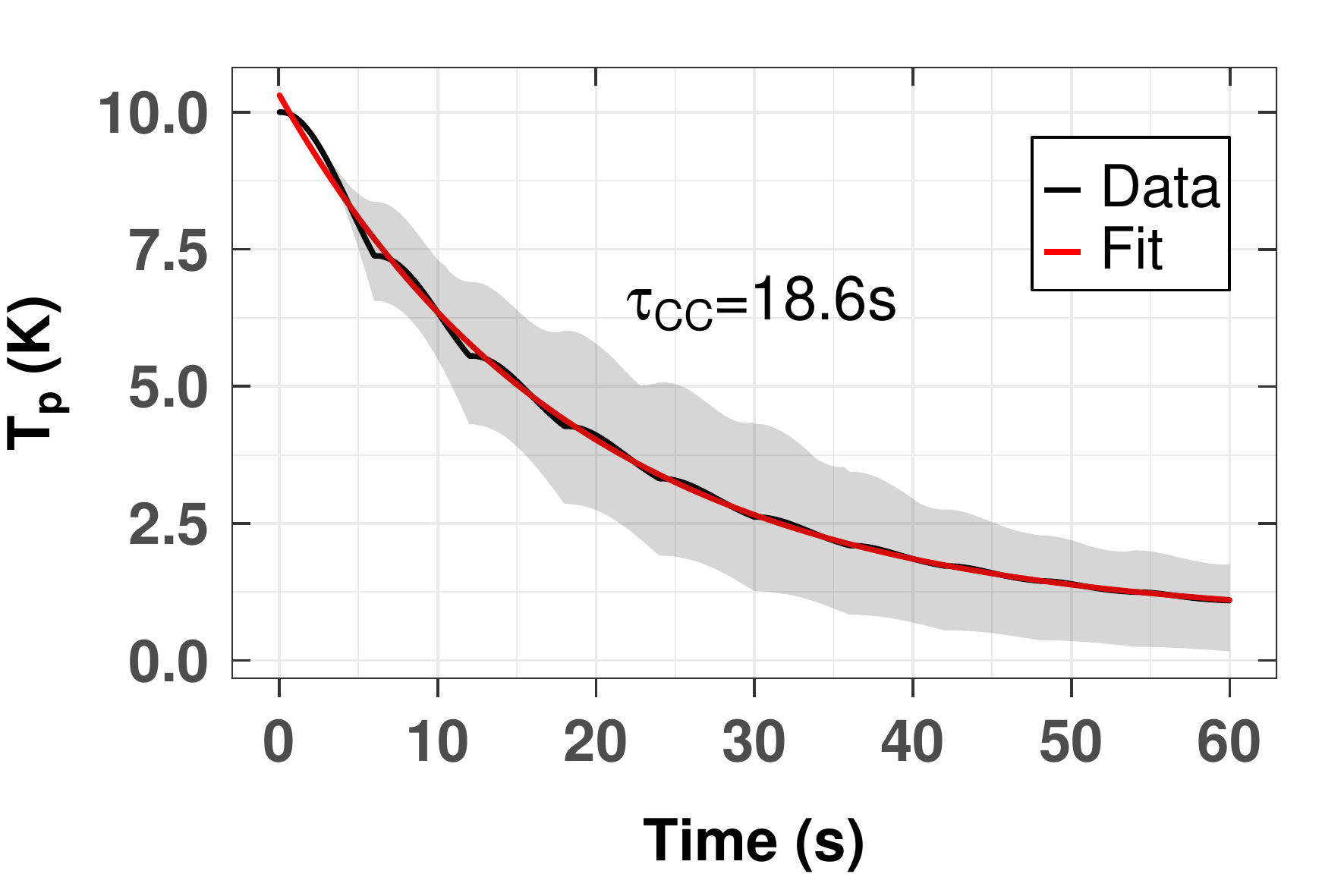}}{ (a) }  & 
			\subf{\includegraphics[width=0.47\linewidth]{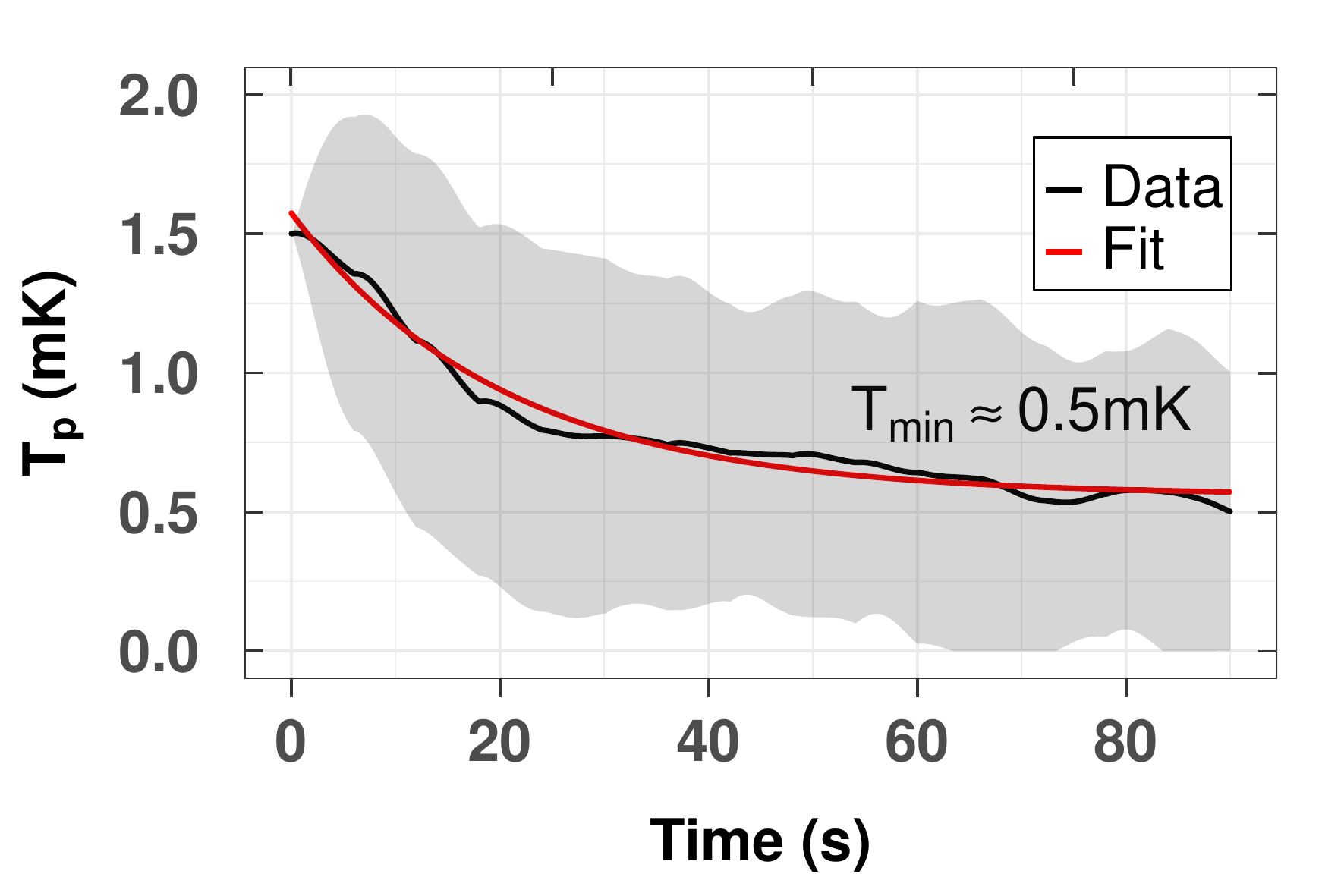}}{ (b) } \\
		\end{tabular}
	}
	\caption{ (a) Simulated proton energy over time when applying the  chopped laser scheme for 80  \Be{} ions, yielding a cooling time constant of $\tau_\text{CC} = \SI{18.6}{s}$. The depicted data is an average temperature for initial axial frequency mismatches of $\Delta_\text{unc} = \SI{100}{mHz}$ and a dynamical stability of $\xi_\text{stab} = \SI{1e-7}{}$ as defined in Sec.~\ref{sec:def_delta_unc}. The gray band indicates one standard deviation.  (b) Same as in (a), but with the proton initialized at 1.5\,mK. Due to the 200\,kHz-detuning from the resonator effectively no heating sources are present, such that the ultimate temperature limit is the Doppler temperature of the beryllium ions of $T_D \approx \SI{0.5}{mK}$. }
	\label{fig:CC_result}
\end{figure*}

 	 To determine a realistic cooling time constant, we incorporate the experimental frequency uncertainties in this  section and the following ones  twofold: First, we add an additional detuning $\delta \nu$ to the proton frequency due to non-ideal matching to the  \Be{} frequency. $\delta \nu$ is drawn uniformly from the interval  $\delta \nu \in [-\Delta_\text{unc}, +\Delta_\text{unc}]$ with $\Delta_\text{unc} = 100$\,mHz and randomly for every simulation. Second, we assume a relative voltage stability as defined in Sec.~\ref{sec:numerical_implementation} of $\xi_\text{stab}=\SI{1e-7}{}$ for both the proton and beryllium trapping voltages.
 	In a first step we determine the ideal laser  cycle time for the chopped laser scheme. 
 	The \Be{} ions are initialized at an energy of $\SI{1}{mK}\cdot k_B$ and the proton at an energy of 10~\kkb. 
 	As figure of merit we  use the average temperature gain of the beryllium cloud over time, $\Delta T_{\text{Be}}/\Delta t$, displayed in Fig.\,\ref{fig:CC_laser_duty}(a). By switching on the cooling laser at the time of maximum $\Delta T_{\text{Be}}/\Delta t$, in this simulated case at  6.0\,s,  to first order the fastest cooling time constant is achieved. We note that more complex laser schemes with varying cycle times could potentially lead to an even better cooling performance, however, in this paper we will limit the discussion to schemes with constant laser cycle times. The on-time of the laser is set to  2\% of the laser  cycle time. Fig.\,\ref{fig:CC_laser_duty}(b) shows a representative cooling run with the cooling laser turned on every  6.0~s. Averaging several of such runs with the uncertainties described above allows us to perform a least-square fit with an exponential function 
 	\begin{equation}
 	T_\text{p} = T_{p,0} \,e^{-t/\tau_\text{CC}} + T_\text{p,min}
 	\label{eq:exponential_fit}
 	\end{equation}
 	 and extract a cooling time constant $\tau_\text{CC}$ as it is done in Fig.\,\ref{fig:CC_result}(a).  We obtain $\tau_\text{CC} = 18.6(1.4)$\,s for the simulated case with 80  \Be{} ions. The uncertainty is estimated by fitting the single runs and calculating the standard deviation of the mean $\tau_\text{CC}$. 
 	
 	In addition to the cooling time constant, the other parameter of interest is the lower temperature limit, determined by the equilibrium of the competing cooling and heating rate. In case of the  common-endcap coupling, the heating by the resonator noise is strongly suppressed  due to the resonator detuning of several hundred kHz. Additionally, other sources of heating are negligibly small \cite{Bor19}. We investigate the lower limit that can be achieved in the simulations by repeating the procedure for the cooling time constant, but initializing the proton at 1.5\,mK, shown in Fig.\,\ref{fig:CC_result}(b). The cooling time constant is within uncertainty indistinguishable from the one with 10\,K initial proton temperature. More importantly, the lower temperature limit is  dictated by the temperature of the  \Be{} cloud, which itself is cooled to the Doppler limit of $T_\text{D} \approx \SI{0.5}{mK} $.

\begin{figure*}
 		\centering
 		\centerline{
 		\begin{tabular}[t]{ll}
 			\rule{0pt}{-1.0ex}%
 			\subf{\includegraphics[width=0.47\linewidth]{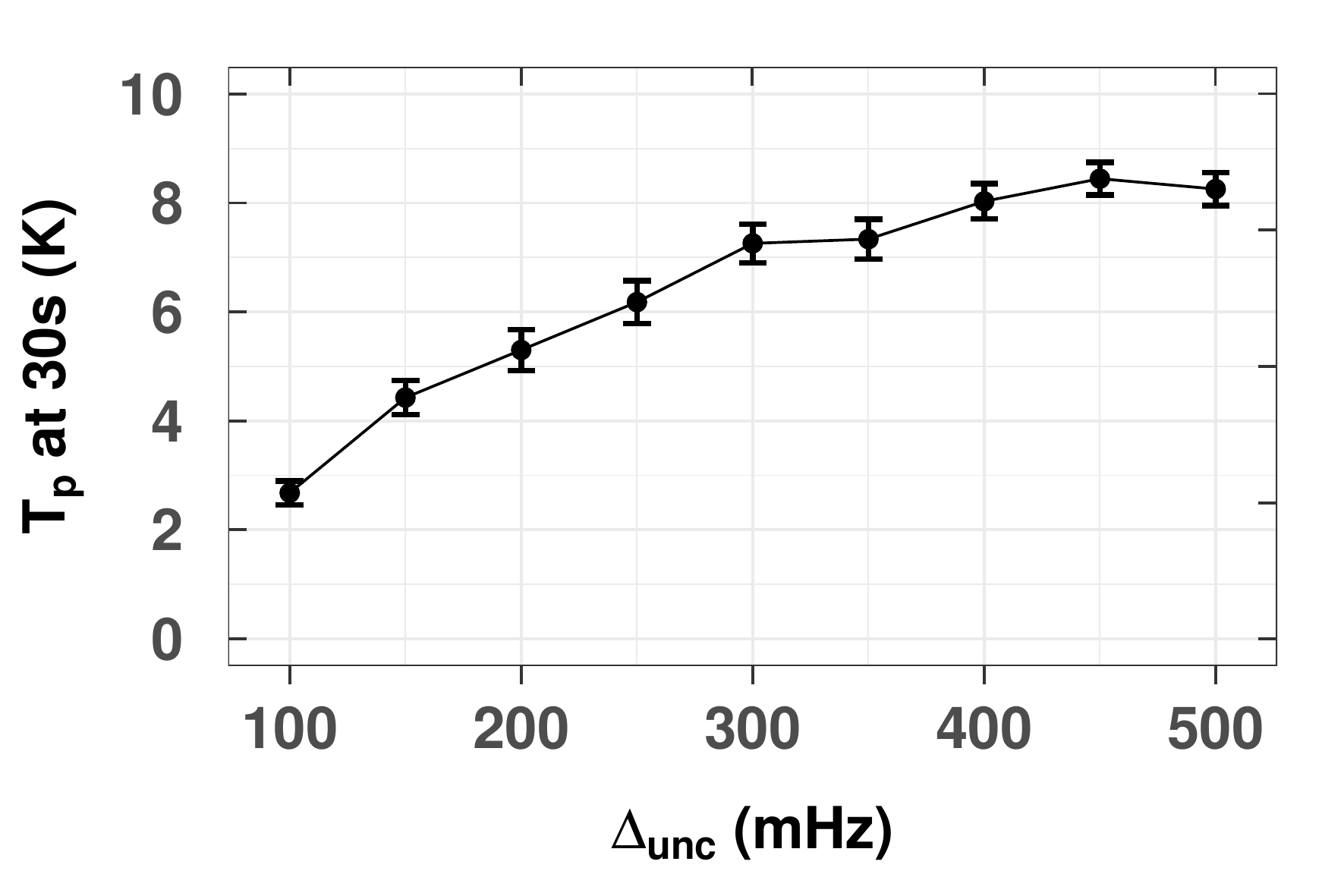}}{ (a) }  &
 			\subf{\includegraphics[width=0.47\linewidth]{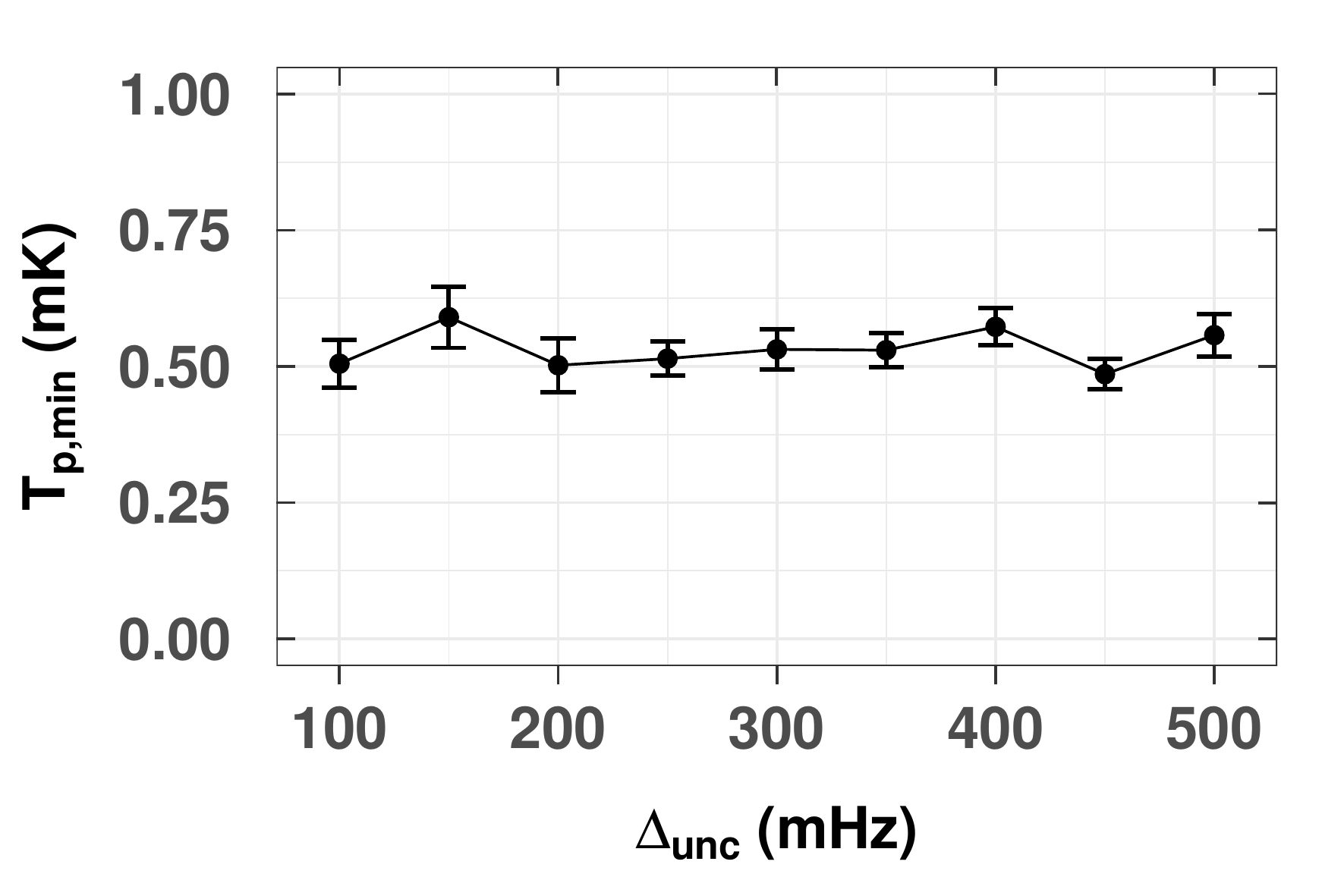}}{ (b) } \\
 		\end{tabular}
 	}
 		\caption{ Simulation of the sensitivity of the  common-endcap coupling approach to relative  frequency matching uncertainties between the proton and 80 \Be{} ions. In (a) the proton temperature at 30\,s after initialization as a measure of the cooling time is shown. 
 			In (b) the lower temperature limit as a function of $\Delta_\text{unc}$ is displayed for a proton initialized at $\SI{0.5}{mK}\cdot k_B$. Due to the absence of heating effects in the  common-endcap coupling scheme the lower temperature limit for the proton is not affected by a detuning and  remains at the Doppler limit of $T_\text{D} \approx \SI{0.5}{mK}$.  }
 		\label{fig:CC_sensitivity_detuning}
 	\end{figure*}

 	The last quantity of practical interest is the sensitivity of the cooling performance to relative axial frequency detunings between the proton and the  \Be{} cloud. The cooling efficiency for a fixed relative detuning depends on the laser  cycle time. In the worst case, even for a rather small relative detuning of $\approx 10$\,mHz no cooling can occur if the laser  cycle time is an integer multiple of the Rabi cycle.
 	 The relevant quantity in the experiment is the precision with which the particle frequencies can be matched. Hence we vary the frequency matching precision $\Delta_\text{unc}$ as defined above instead of introducing a fixed detuning. 
 	Furthermore, since for large detuning intervals fitting the temperature evolution with Eq.\,(\ref{eq:exponential_fit}) becomes an erroneous measure of cooling effectiveness, we will use the mean proton temperature at 30\,s after initialization instead. 
 	Fig.\,\ref{fig:CC_sensitivity_detuning}(a) depicts the sensitivity of the proton temperature at 30\,s as a function of the detuning uncertainty. The initial proton energy was 10~\kkb. 
 The proton temperature at 30\,s nearly doubles when moving from $100$\,mHz to $200$\,mHz, indicating that absolute detunings of $> 100$\,mHz contribute little to the cooling process, which is a result of the comparably low Rabi frequency of \SI{34}{mHz}.
  Fig.\,\ref{fig:CC_sensitivity_detuning}(b) shows  the simulated proton temperature after initializing the proton at 0.5\,mK$\cdot k_B$ and letting it evolve for 30\,s while applying the chopped laser  to the  \Be{} ions.  Since for the  common-endcap coupling the resonator is detuned, no heating process competes with the cooling process. As a result, the lower temperature limit is not affected and the proton temperature can still reach the Doppler limit of $T_\text{D} \approx \SI{0.5}{mK}$.

	\subsection{Resonant coupling mediated by an RLC resonator}
	\label{sec:resonant_coupling}
	
	A sketch of the experimental setup for the resonant coupling scheme is shown in Fig.\,\ref{fig:experimental_setup_together}(b) and the experimental proof of principle was recently demonstrated in Ref.\,\cite{Boh21}. This approach is the easiest to observe experimentally as the proton and the  \Be{} ions are continuously being detected in resonance with the RLC resonator, which is the standard operation mode of our Penning-trap experiment.

	In the case of the resonant coupling with an RLC circuit, the resonator noise causes the energy exchange between the proton and \Be{} ions to be highly incoherent on a ms-timescale.	As a result, a cooling scheme that relies on single $\pi$-pulses will not be able to consistently achieve the desired 10~mK. Although a chopped laser scheme  might offer an improved cooling performance compared to a continuous laser operation, we will investigate the continuous case in this work since it has already been experimentally demonstrated \cite{Boh21}.

	As an introductive theoretical analysis, we first consider two identical particles in two separate but identical traps connected to the same resonator. The motion of these two particles can be described by a symmetric and antisymmetric mode, where the individual oscillations are in phase or shifted by a phase of $\pi$, respectively. 
	The individual image currents induced into the resonator interfere destructively for the antisymmetric mode. As a result, the resonator only couples to the symmetric mode and only this mode is thermalized to the detector temperature. The same argument applies to two identical ions in the same trap.
	The experimental evidence for this is provided by the linear scaling of the dip width with the number of ions in the trap, which arises from substituting $q \rightarrow N_i \cdot q$ and $m \rightarrow N_i \cdot m$ in Eq.\,(\ref{eq:dipwidth}). However, this substitution is only valid if the particles' motions are in phase. On the other hand, by introducing an additional damping constant to one of the particles, energy is transferred between the symmetric and antisymmetric mode. In our experiment, the cooling laser constitutes this additional damping constant. 
	The fact that only the symmetric mode is heated by the resonator and the damping of the  \Be{} ions by the laser continuously transfers energy between all modes implies that even a weak intensity laser is sufficient to eventually cool all modes except the symmetric one down to the Doppler limit. In the case of negligible energy of the non-symmetric modes, the final proton temperature is determined by the proton's relative component of the symmetric mode temperature. The relative components are given by the  relative dip widths, $\gamma_\text{p}$ and $\gamma_\text{Be}$, respectively:
	\begin{align}
		T_\text{res} = T_\text{p} + T_\text{Be} \\
		T_\text{p} = \frac{ \gamma_\text{p} }{ \gamma_\text{Be} } T_\text{Be}.
	\end{align}
	These equations result in the $T_p \propto 1/N_\text{Be}$-scaling initially proposed in Ref.\,\cite{Hei90} and measured in Ref.\,\cite{Boh21}:
	\begin{align}
		T_\text{p} = \frac{1}{1 + \frac{ \gamma_\text{Be}}{ \gamma_\text{p} }} T_\text{res} \appropto \frac{1}{N_\text{Be}}.
		\label{eq:Tpgamma}
	\end{align}
	However, in this form Eq.\,(\ref{eq:Tpgamma}) is only true for weak laser intensities, i.e. if the  damping rate of the \Be{} ions due to the laser, $\gamma_\text{L}$, is much smaller than $\gamma_\text{Be}$. At higher laser intensities, the beryllium cloud is decoupled from the resonator.  This effect can be accounted for by a reduction of the effective dip width, expressed by the substitution $\gamma_{\text{Be}} \rightarrow \tilde \gamma_{\text{Be}} = k \,\gamma_{\text{Be}}$ with dimensionless scaling parameter $0 \leq k \leq 1$ in Ref.\,\cite{Boh21}.

\begin{figure*}
			\centering
			\centerline{
			\begin{tabular}[t]{ll}
				\rule{0pt}{-1.0ex}%
				\subf{\includegraphics[width=0.48\linewidth]{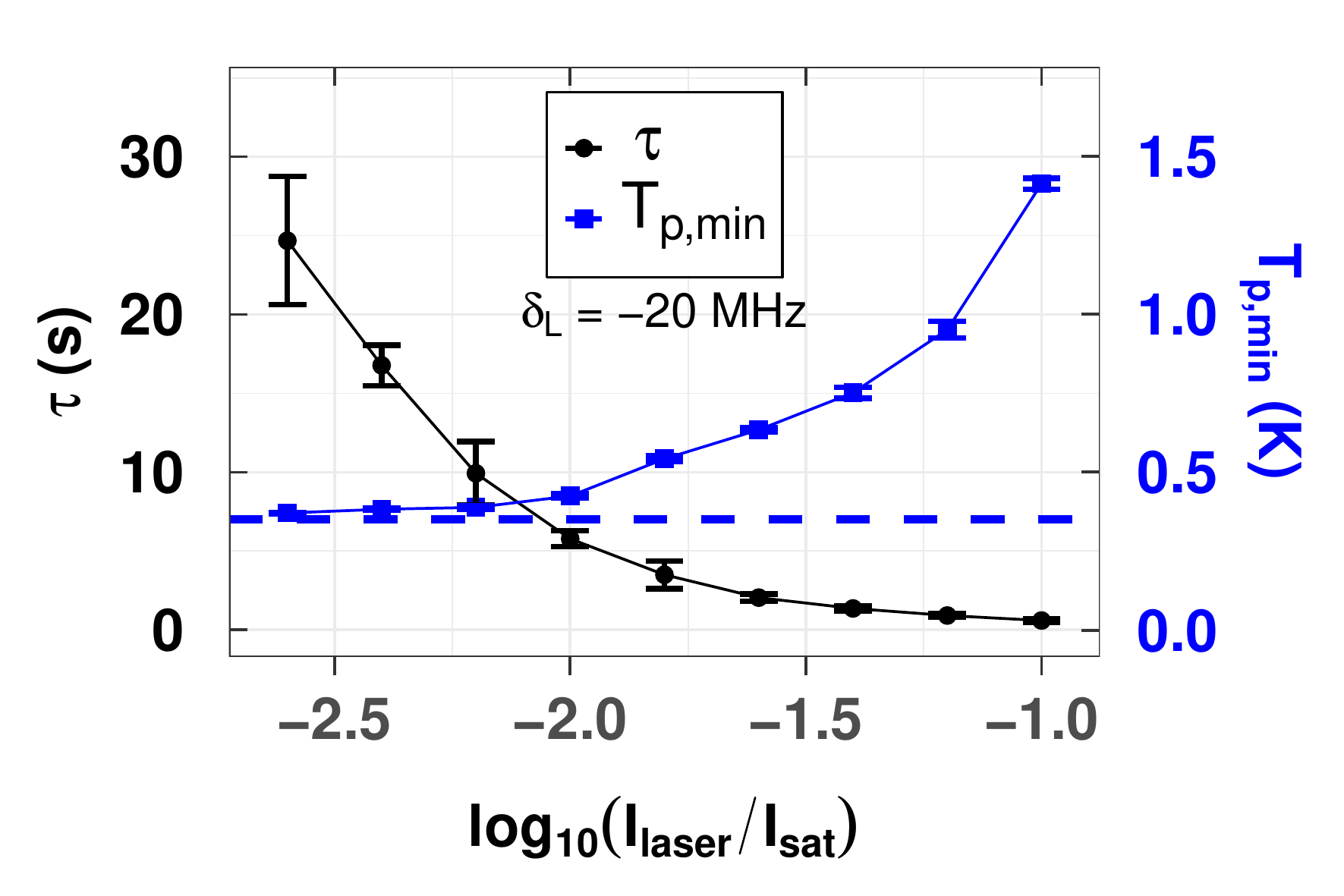}}{(a) }  &
				\subf{\includegraphics[width=0.48\linewidth]{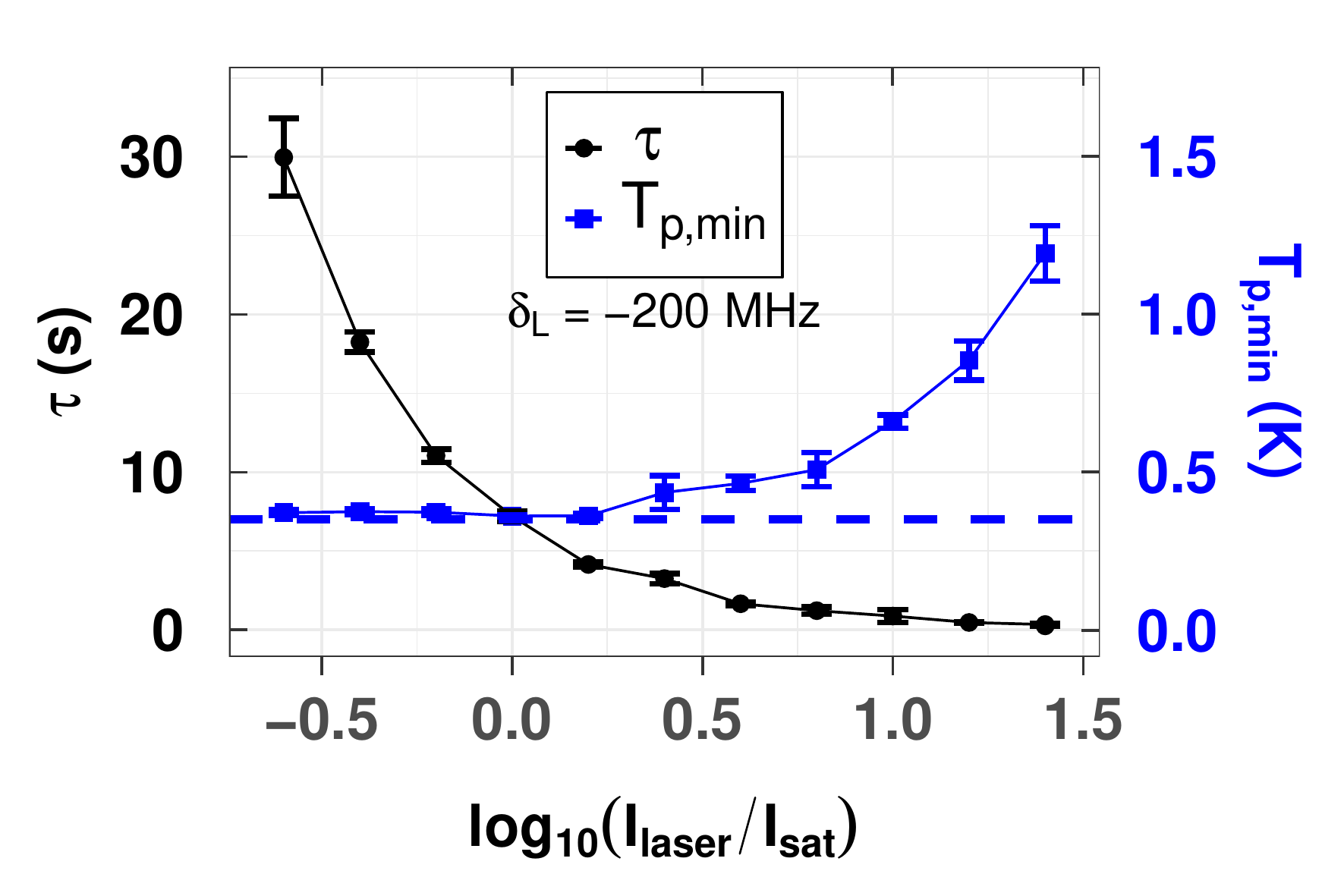}}{ (b) } 
%				\subf{\includegraphics[width=0.47\linewidth]{RC_tau_lowerLimit_combined_100MHz.png}}{ (c) }  &
%				\subf{\includegraphics[width=0.47\linewidth]{RC_tau_lowerLimit_combined_200MHz.png}}{ d) } \\
			\end{tabular}
		}
			\caption{Simulated cooling time constant and lower proton temperature limit as a function of laser intensity for laser detunings $\delta_L = -20$\,MHz  (a) and $\delta_L= -200$\,MHz  (b). Although the absolute values of the laser intensity strongly differ, the scaling of $\tau$ and $T_\text{p,min}$ is very similar, indicating that no global favourable laser detuning exists and rather the combination of laser intensity and detuning defines the cooling performance. The blue dashed lines indicate the minimal proton temperature of 350\,mK predicted by Eq.\,(\ref{eq:Tpgamma}).}
			\label{fig:RC_tau_lowerLimit_combined}
		\end{figure*}

	Our simulation results show that,  in contrast to the  lower limit of the proton temperature, a higher laser intensity is beneficial for the cooling time constant as it increases the rate of energy dissipation. In order to study these relations quantitatively in terms of experimental parameters, we simulate the proton and a cloud of 80 continuously  laser-cooled  \Be{} ions tuned on resonance with the detector.  
	Fig.\,\ref{fig:RC_tau_lowerLimit_combined} shows  the cooling time constant $\tau_\text{RC}$ and the lower temperature limit $T_\text{p,min}$ as a function of laser intensity for two exemplary laser detunings of $-20$\,MHz and $-200$\,MHz, depicted in subfigures (a) and (b), respectively. The cooling time constant is determined by least-squares fits with Eq.\,(\ref{eq:exponential_fit}) for an initial proton energy of 10~\kkb. The lower temperature limit is determined by initializing the proton at 0~\kkb{} and averaging the proton energy after it has reached equilibrium.
	
	While a much stronger laser intensity is required for $-200$\,MHz laser detuning, the general behavior of $\tau$ and $T_\text{p,min}$ are the same:
	The cooling time constant drops with larger laser intensity as the  \Be{} ions are damped more strongly and reaches values down to a few hundred ms. 
	However, at such high laser intensities the beryllium ions are decoupled too strongly from the resonator, leading to a suboptimal proton temperature limit. Instead, the proton temperature saturates in the region where $\gamma_\text{L} \ll \gamma_\text{Be}$. The simulated values in the weak intensity limit agree well with the theoretical prediction of $T_\text{p,min} = 350$\,mK by Eq.\,(\ref{eq:Tpgamma}), indicated by the blue dashed line.

\begin{figure*}
	\centering
	\centerline{
	\begin{tabular}[t]{ll}
		\rule{0pt}{-1.0ex}%
		\subf{\includegraphics[width=0.48\linewidth]{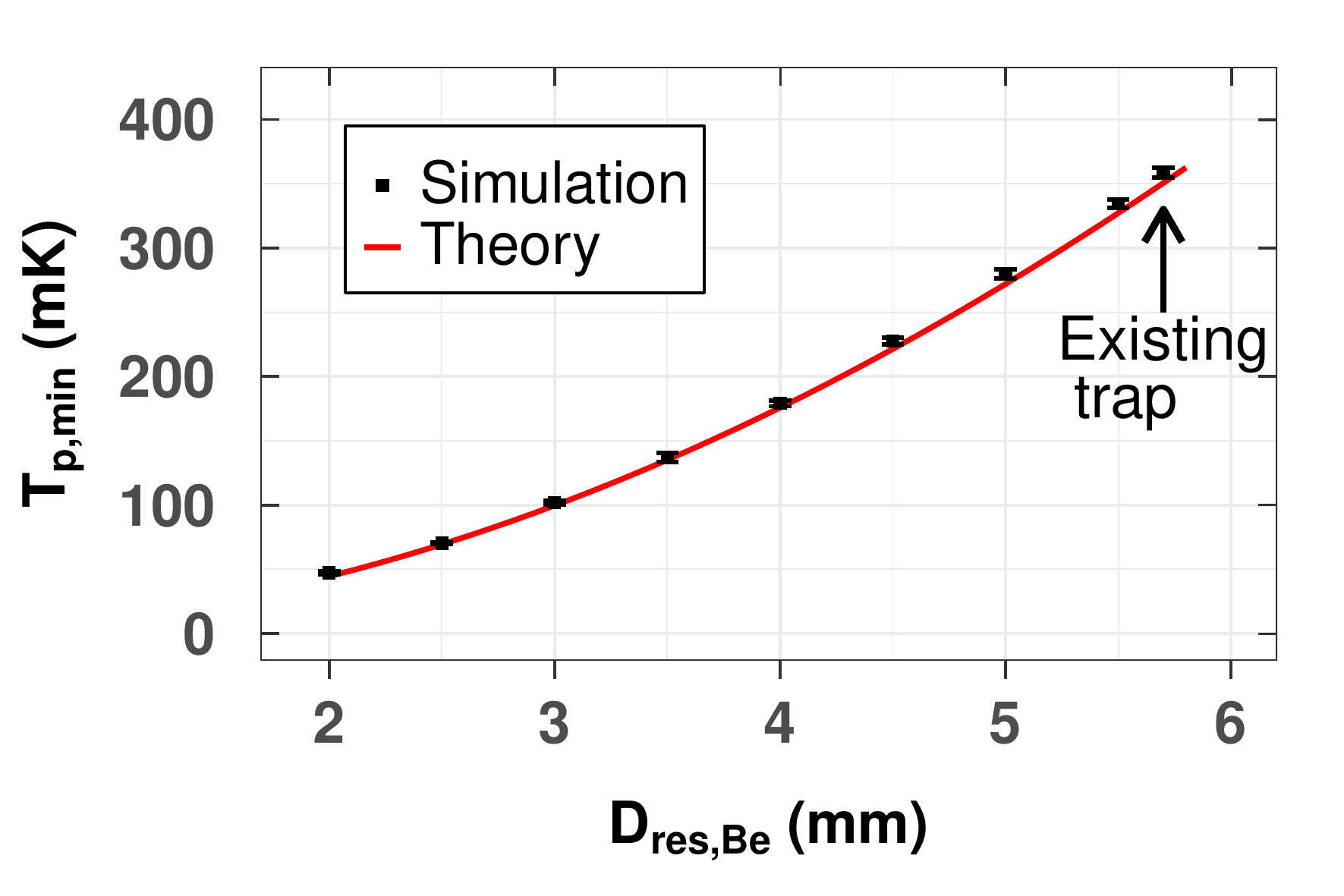}}{(a) }  &
		\subf{\includegraphics[width=0.48\linewidth]{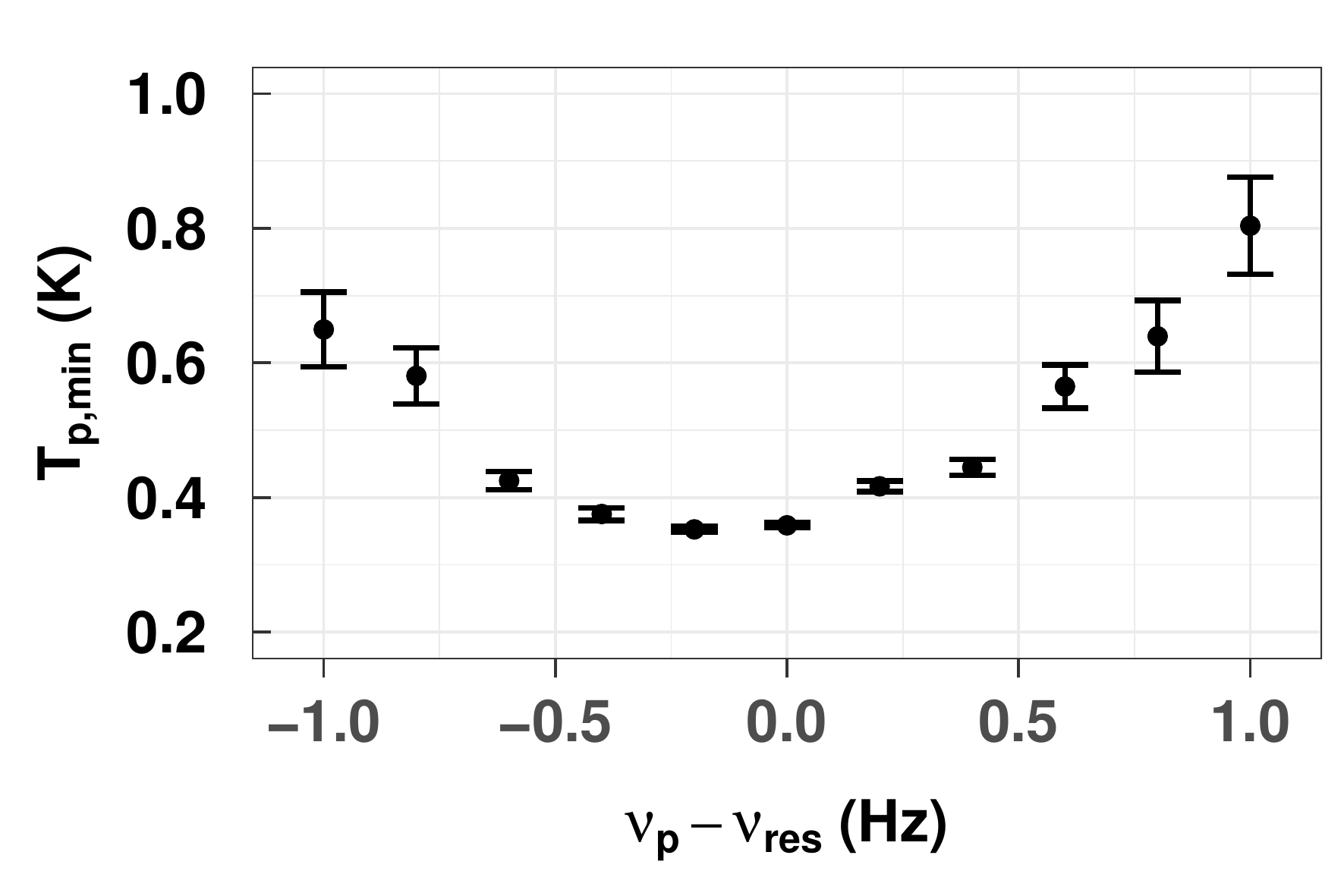}}{ (b) } \\
	\end{tabular}
	}
	\caption{(a) The  minimal proton temperature as a function of the effective beryllium trap size for 80  \Be{} ions. A smaller trap size increases the  width of the beryllium dip, leading to a lower  minimal proton temperature according to Eq.~(\ref{eq:Tpgamma}). The simulation data is generated with  sufficiently low laser power. (b) The simulated final proton temperature as a function of the detuning of the  proton from the resonator. The asymmetry arises from the effect described in Sec.\,\ref{sec:capacitive_coupling}. 
 		}
	\label{fig:RC_trap_size_and_detuning}
\end{figure*}

	According to  Eq.\,(\ref{eq:dipwidth}), the beryllium dip width can be increased not only by loading more beryllium ions, but also by reducing the  effective electrode distance $D_\text{res}$.
	Since the cooling time constants are a function of both the laser intensity and $D_\text{res,Be}$, we use $T_\text{p,min}$ in the weak intensity limit as a measure of improved cooling performance.
	In Fig.\,\ref{fig:RC_trap_size_and_detuning}(a), $D_\text{res,Be}$ is decreased starting from the currently manufactured trap with $D_\text{res,Be} = \SI{5.7}{mm}$. 
	 The red line is the prediction of Eq.~(\ref{eq:Tpgamma}) and we observe excellent agreement between simulation and theory.
	 However, even for a trap with $D_\text{res,Be} = \SI{2.0}{mm}$, which is  challenging to manufacture and difficult to operate with 80  \Be{} ions, the lower temperature limit is only about 50\,mK, a factor of 5 higher than our desired temperature.

	While the resonant coupling scheme with continuous laser operation has a comparably high temperature limit, it is the most robust technique against experimental uncertainties. Fig.\,\ref{fig:RC_trap_size_and_detuning}(b) shows the final proton temperature as a function of the detuning of the  \Be{} cloud. 
	The fact that the minimal $T_\text{p,min}$ is shifted from 0\,Hz detuning is caused by the effect of maximizing the coupling strength between the two particle species by introducing a relative detuning as described in Sec.\,\ref{sec:capacitive_coupling}. However, since the laser is not  chopped here, the effect on the cooling performance is negligible.
	Since the particles are continuously  detected on resonance, matching and stabilizing them to $\approx 100$\,mHz is comparably straightforward and routinely done in our experiment. As Fig.\,\ref{fig:RC_trap_size_and_detuning}(b) shows, the increase in $T_\text{p,min}$ is negligible for $\approx$ 100\,mHz detuning. 
	
	\FloatBarrier
	
	\subsection{Off-resonant coupling mediated by an RLC resonator}
	\label{sec:off_resonant_coupling}

	The off-resonant resonator coupling approach combines the ideas of the previous two cooling schemes by detuning the particle frequencies by $\delta \nu_\text{res}$ from the resonator eigenfrequency, typically several kHz \cite{Tu21}. 	Similar to the  common-endcap coupling, the particles then perform Rabi oscillations mediated by a common effective capacitance
	\begin{equation}
		C_\text{eff} = \frac{\Im(Z_\text{RLC}^{-1})}{\omega} \appropto \delta \nu_\text{res}
		\label{eq:C_eff}
	\end{equation}
	where $Z_\text{RLC}$ is the impedance of the RLC detection circuit. Combined with Eq.\,(\ref{eq:CC_rabi}) this implies that the Rabi frequency decreases approximately linear with the resonator detuning. With this scheme, the application of $\pi$-pulses is not only inhibited by the required frequency matching precision, but also by the residual heating due to the resonator noise. As a result, we employ a chopped laser scheme again.
	
\begin{figure*}
		\centering
		\centerline{
		\begin{tabular}[t]{ll}
			\rule{0pt}{-1.0ex}%
			\subf{\includegraphics[width=0.47\linewidth]{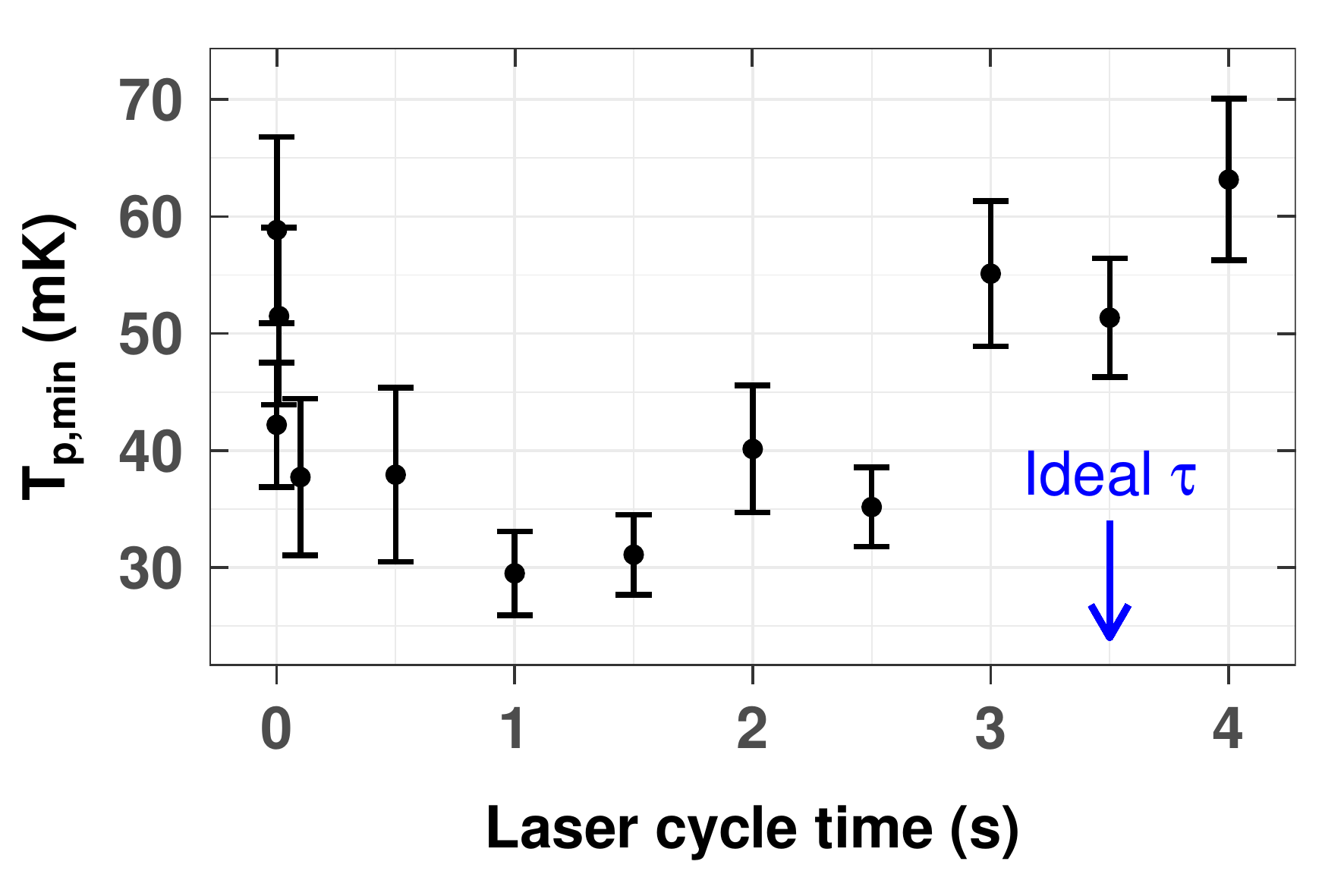}}{(a) }  &
			\subf{\includegraphics[width=0.47\linewidth]{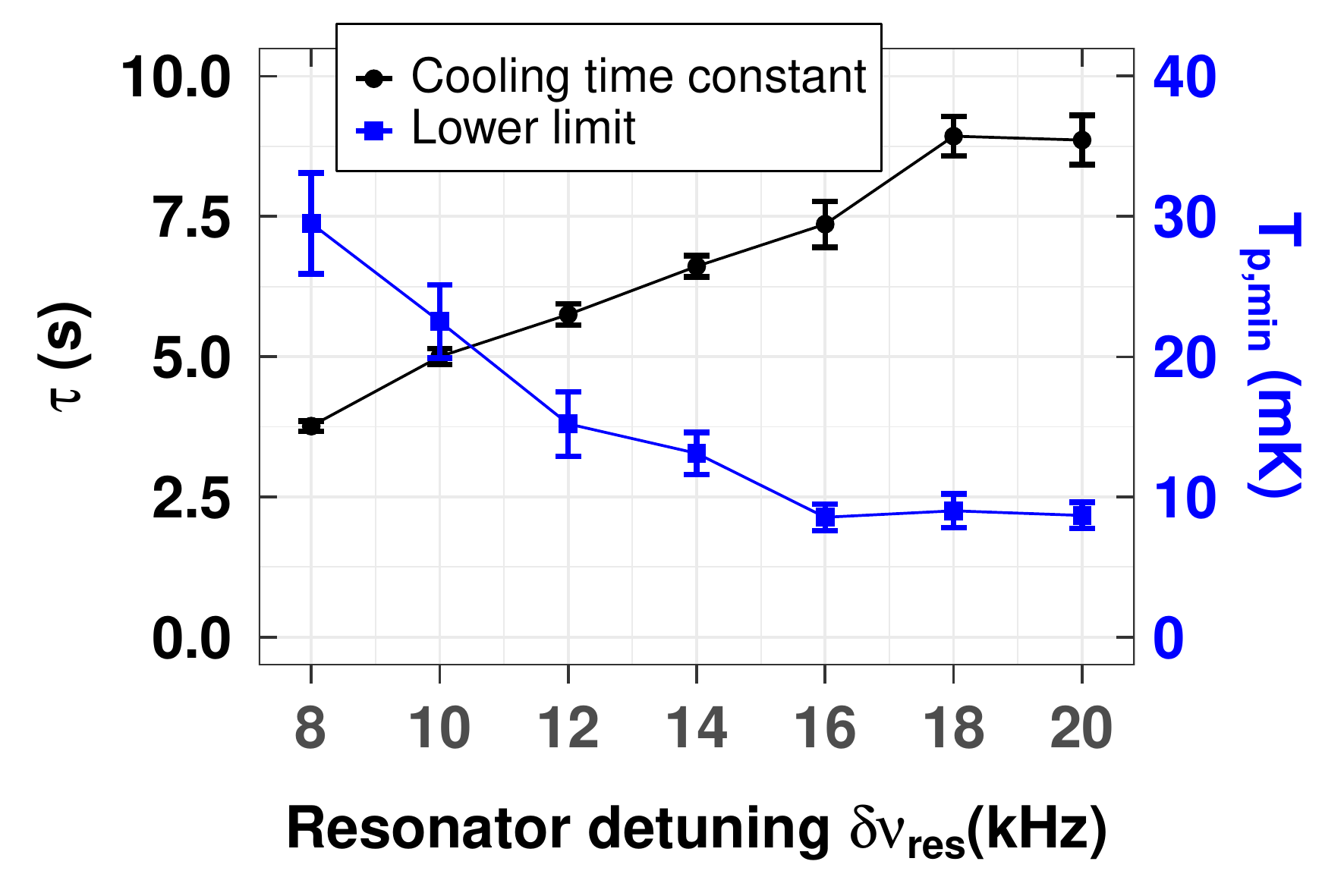}}{ (b) } \\
		\end{tabular}
	}
		\caption{(a) The simulated proton temperature limit as a function of laser  cycle time for 8\,kHz resonator detuning. Due to the resonator noise the ideal laser  cycle time shifts from 3.5\,s for the cooling time constant to 1.0\,s for the minimal temperature.  The behavior is similar for all resonator detunings. 
			(b) Simulated cooling time constant and temperature limit as derived from (a) as a function of resonator detuning. }
		\label{fig:HC_tau_lowerLimit}
	\end{figure*}

	In order to pinpoint the optimal laser  cycle time and cooling time constant, we follow  the same procedure as for the  common-endcap coupling for different resonator detunings. 
	However, due to the residual particle heating by the resonator noise, the lower temperature limit is higher than the Doppler limit and is a function of $\delta \nu_\text{res}$. Moreover, the incoherence introduced by the noise causes the ideal laser  cycle time for the cooling time constant to become suboptimal for reaching the lowest final temperatures. Fig.\,\ref{fig:HC_tau_lowerLimit}(a) shows the simulated proton temperature as a function of laser  cycle time at 8\,kHz resonator detuning. 
	The proton was initialized at 50\,mK and the system was evolved for 30\,s to allow it to come into equilibrium. The displayed temperature is  calculated from the mean energy of an additional 30\,s-interval afterwards. 
	While the laser  cycle time for the cooling time constant is optimal at 3.5\,s,  about a factor of 2 in the lower temperature limit is gained by reducing it to 1.0\,s.  Equivalent scans were repeated for each  resonator detuning in the following data. 
	
	The simulated performance of the off-resonant coupling scheme in terms of cooling time constants and lower temperature limits is summarized in Fig.\,\ref{fig:HC_tau_lowerLimit}(b). The cooling time constants rise roughly linearly with the detuning, as  expected from  Eq.\,(\ref{eq:CC_rabi}) and Eq.\,(\ref{eq:C_eff}). The lower temperature limits saturate at slightly below 10\,mK, which results from the two opposing effects of the decline of resonator noise and the reduction of the Rabi frequency. 
	
\begin{figure*}
		\centering
		\centerline{
		\begin{tabular}[t]{ll}
			\rule{0pt}{-1.0ex}%
			\subf{\includegraphics[width=0.47\linewidth]{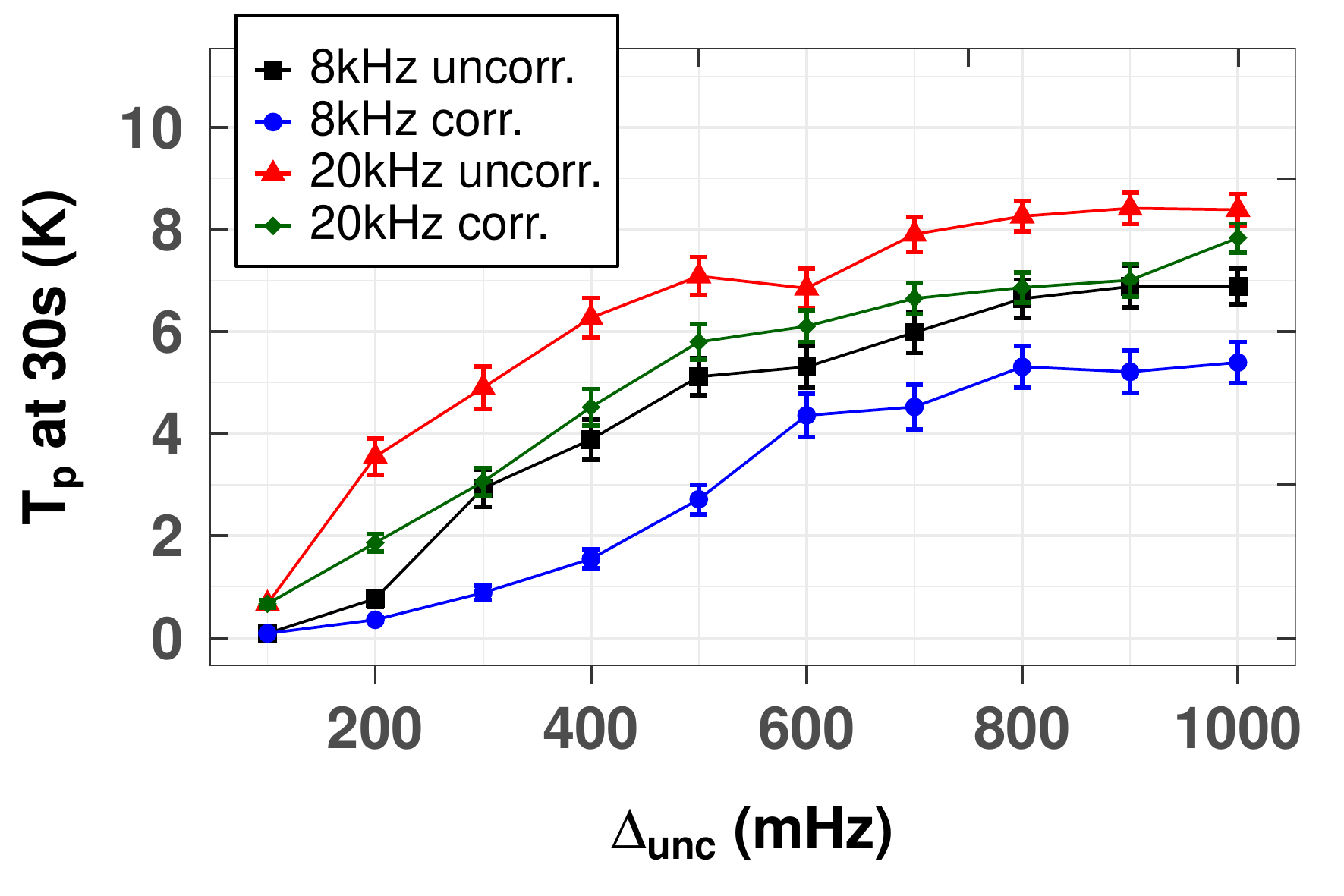}}{ (a) }  &
			\subf{\includegraphics[width=0.47\linewidth]{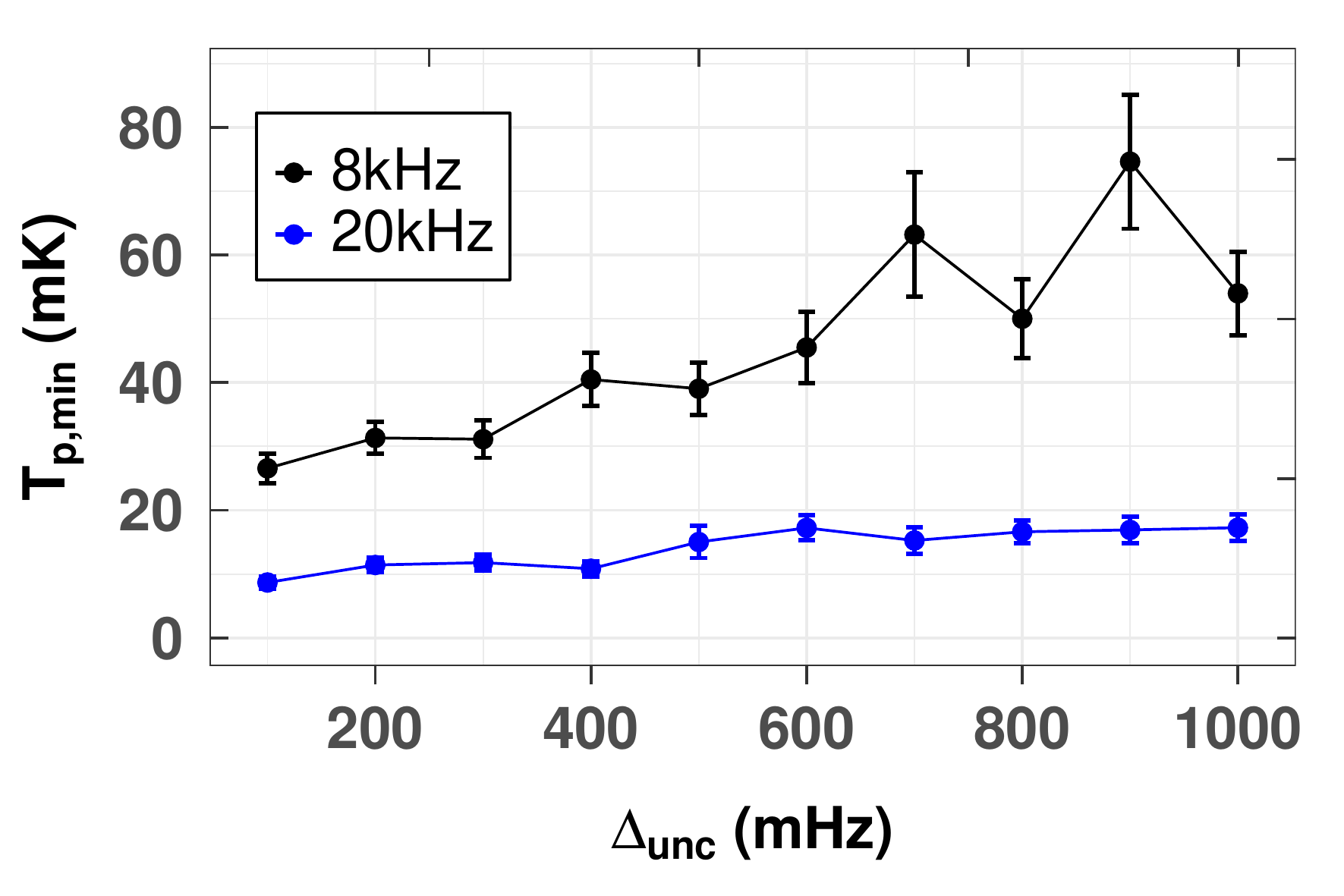}}{ (b) } \\
		\end{tabular}
	}
		\caption{The simulated effect of the frequency matching uncertainty $\Delta_\text{unc}$ for the off-resonant coupling scheme for  resonator detunings of 8\,kHz and 20\,kHz. The data points labelled ``uncorr.'' (uncorrected) were generated with the ideal laser cycle time for $\Delta_\text{unc} = 100$\,mHz. The data points labelled  ``corr.'' (for corrected) were generated with the ideal laser cycle time for each value of $\Delta_\text{unc}$.
		(a) The proton temperature at 30\,s as a measure of cooling effectiveness. The initial energy was 10~\kkb. (b) The lower limit of the proton temperature with the same uncertainties as in (a).   }
		\label{fig:HC_detuning_sensitivity}
	\end{figure*}

	We  investigate the effect of the frequency matching uncertainty for 8\,kHz and 20\,kHz resonator detuning in the same manner as for the  common-endcap coupling. Fig.\,\ref{fig:HC_detuning_sensitivity}(a) shows the  temperature at 30\,s after initializing the proton at an energy of 10\,\kkb. 
	If the particle frequency matching precision $\Delta_\text{unc}$  is known experimentally, one can optimize the cooling time by adjusting the laser cycle time accordingly. For example, if the frequency matching precision is worse than the $100$\,mHz assumed in this work, one can decrease the laser cycle time to account for the faster (but incomplete) Rabi oscillations.  However, this requires $\Delta_\text{unc}$ to be smaller than or at least comparable to the Rabi frequency. This condition is, in contrast to the common-endcap coupling, fulfilled for the coupling scheme investigated in this section. 
	The procedure in context of Fig.\,\ref{fig:CC_laser_duty}(a) was repeated for every frequency matching precision for the blue and green data points labelled ``corr.'' in Fig.\,\ref{fig:HC_detuning_sensitivity}(a), leading to a significantly better cooling performance compared to the uncorrected ones. 
	Especially  in the case of 8\,kHz the sensitivity of the proton temperature to $\Delta_\text{unc}$ is significantly smaller compared to the  common-endcap coupling.  Additionally, Fig.\,\ref{fig:HC_detuning_sensitivity}(b) shows the equilibrium proton temperature as a function of the frequency matching uncertainty. An uncertainty of 1\,Hz causes the temperature to be roughly a factor of 2 higher than for 100\,mHz for both 8\,kHz and 20\,kHz resonator detuning. 

	\FloatBarrier
	
	\section{Conclusion and Outlook}

	We have theoretically studied the sympathetic cooling of the axial mode of a single proton with  laser-cooled  \Be{} ions. A summary of the results is given in Tab.\,1. Of the three studied cooling techniques, the  common-endcap coupling scheme has a comparably long  cooling time constant, but still outperforms the established method of resistive cooling. Due to the absence of resonator heating, the proton temperature is only limited by the Doppler limit of the beryllium ions. However, the method is experimentally challenging since the particles cannot be detected during the coupling process. Additionally, the  low Rabi frequency of $\approx 30$\,mHz makes it the  most sensitive scheme to relative frequency mismatches between the particles. In contrast, the RLC resonator mediated resonant coupling scheme can be tuned to have cooling time constants in the $100$\,ms-regime. It requires no additional experimental instrumentation to conduct since all three oscillators are being detected on resonance, which is the standard operation mode of our Penning trap. However, it is limited to $\approx 350$\,mK axial proton temperature with the current experimental setup. Nevertheless, it will certainly be used as a pre-cooling mechanism or as an alternative to feedback cooling. The off-resonant RLC resonator coupling technique combines both approaches and yields a fast cooling time constant while capable of achieving $\approx 10$\,mK proton temperature, which suffices as cooling scheme to prepare cold (anti-)protons for $g$-factor measurements \cite{Boh17, Smo17, Moo13}. Furthermore, it is more robust than the  common-endcap coupling against relative frequency mismatches. Its main advantage is that the ratio between cooling time constant and lower temperature limit can be adjusted to the needs of the  particular application.
	
	\renewcommand\arraystretch{1.6}
	\setlength\tabcolsep{3.0pt}

		\begin{table}
				\begin{center}
		\begin{tabular}{c|c|c|c}
			Method  & $\tau (s)$ & $T_\text{p,min}$ (K) & Sensitivity to detuning \\
			\hline 
			Common-endcap coupling & 20\,s & 0.5\,mK & off-resonance, $\approx 200$\,mHz \\
			Resonant RLC coupling & 0.1\,--\,30\,s & 350\,mK & on-resonance, $\approx 1$\,Hz \\
			Off-resonant RLC coupling & 4\,--\,10\,s & 10\,--\,30\,mK & off-resonance,  $\approx 200$ -- $500$\,mHz
			\label{tab:summary}
		\end{tabular}
			\caption{Summary of the three different coupling schemes in terms of the cooling time constant $\tau$ and the proton temperature limit $T_\text{p,min}$. 
				The sensitivity to detuning is defined as the frequency matching precision where the cooling performance is a factor of 2 worse than the reference at 100\,mHz. }
				\end{center}
		\end{table}

	Overall, the studies presented here establish the cooling methods for the next generation of precision measurements at BASE. While the simulations already supported the proof-of-principle measurement in Ref.~\cite{Boh21}, they allow  a quantitative assessment of the performance of  future techniques and  an identification of their experimental challenges, which would otherwise require dedicated measurement campaigns. 
	The next step is to apply these schemes experimentally and validate the conclusions regarding temperature limits and cooling times drawn in this  work. For this purpose an improved experimental setup is currently being commissioned \cite{WieTh}. It extends the current version by a  dedicated trap for temperature measurements with $\approx 1$\,mK resolution and a cryogenic switch to detune the resonator for the application of the  chopped laser schemes.  
	
	Once cooling to 10\,mK is routinely achieved experimentally, the cooling techniques described here will allow magnetic moment measurements to be conducted at improved sampling rates and reduced systematic shifts and will thus have considerable impact on future tests of CPT invariance  with protons and antiprotons. 
	In addition, the two-trap sympathetic cooling techniques are currently being developed for tests of quantum electrodynamics with  highly charged ions \cite{Stu19} or the determination of the magnetic moment of $^3$He$^{2+}$ \cite{Moo18}.

	 Since the techniques are practically independent of the target ion and the species can be macroscopically separated, applications beyond these specific cases are conceivable. For example, high-precision mass measurements of stable ions \cite{Mye13} or radionuclides with moderate lifetimes at online facilities \cite{Bla06, Ero13}, studies with molecular ions in radiofrequency traps \cite{Cai17} as well as trapped-ion quantum computing \cite{Bru19} could benefit from the techniques presented in this  paper. Moreover, sub-Doppler cooling techniques such as sideband cooling \cite{Goo16, Hrm19} may be employed in the future to reach even lower temperatures.

	\newpage

	\section*{Acknowledgments}
	
	This paper comprises parts of the PhD thesis work of C. Will at the Ruprecht Karl University of Heidelberg.
		
	We thank the \texttt{ALPHATRAP}-group at the Max-Planck-Institute for Nuclear Physics for  helpful discussions regarding 
	the cooling method presented here and acknowledge similar developments toward cooling 
	highly charged ions in the ALPHATRAP collaboration.

	We acknowledge financial support from the Max-Planck-Society, RIKEN Chief Scientist Program, RIKEN
	Pioneering Project Funding, the RIKEN JRA Program, the
	Helmholtz-Gemeinschaft, the DFG through SFB 1227
	‘DQ-mat’, the European Union (Marie Sklodowska-Curie
	grant agreement number 721559), the European Research
	Council (ERC) under the European Union’s Horizon
	2020 research and innovation programme (Grant agree-
	ment Nos. 832848 - FunI and 852818 - STEP), the DAAD RISE program and the
	Max-Planck–RIKEN–PTB Center for Time, Constants
	and Fundamental Symmetries.

	\section*{Competing interests}
	The authors declare no competing interests.

	\bibliography{references} 
	\bibliographystyle{ieeetr}
%	\printbibliography

	\newpage

\end{document}